\documentclass[twocolumn,english,nobibnotes,nofootinbib]{revtex4-1}
\usepackage[T1]{fontenc}
\usepackage[latin9]{inputenc}
\usepackage[a4paper]{geometry}
\usepackage{float}
\usepackage{textcomp}
\usepackage{amsmath}
\usepackage{amssymb}
\usepackage{graphicx}
\usepackage{esint}

\makeatletter
\@ifundefined{textcolor}{}
{%
 \definecolor{BLACK}{gray}{0}
 \definecolor{WHITE}{gray}{1}
 \definecolor{RED}{rgb}{1,0,0}
 \definecolor{GREEN}{rgb}{0,1,0}
 \definecolor{BLUE}{rgb}{0,0,1}
 \definecolor{CYAN}{cmyk}{1,0,0,0}
 \definecolor{MAGENTA}{cmyk}{0,1,0,0}
 \definecolor{YELLOW}{cmyk}{0,0,1,0}
}

\usepackage{babel}

\makeatother

\usepackage{babel}
\begin{document}

\author{K. Kolasi\'{n}ski}

\affiliation{AGH University of Science and Technology, Faculty of Physics and
Applied Computer Science,\\
 al. Mickiewicza 30, 30-059 Kraków, Poland}

\author{A. Mre\'{n}ca-Kolasi\'{n}ska}

\affiliation{AGH University of Science and Technology, Faculty of Physics and
Applied Computer Science,\\
 al. Mickiewicza 30, 30-059 Kraków, Poland}

\author{B. Szafran}

\affiliation{AGH University of Science and Technology, Faculty of Physics and
Applied Computer Science,\\
 al. Mickiewicza 30, 30-059 Kraków, Poland}

\title{
Transconductance and effective Landé factors for quantum point contacts: spin-orbit coupling and interaction effects}
\begin{abstract}

We analyze the effective $g^*$ factors and their dependence on the orientation of the external magnetic field for a quantum point contact
defined in the two-dimensional electron gas. The paper simulates the experimental procedure for evaluation of the effective Land\'e factors
from the transconductance of a biased device in external magnetic field. The contributions of the orbital effects of the magnetic field,
the electron-electron interaction and spin-orbit (SO) coupling are studied.
The anisotropy of the $g^*$ factors for the in-plane magnetic field orientation, which seems counterintuitive from the perspective of the effective SO magnetic field,
is explained in an analytical model of the constriction as due to the SO-induced subband mixing.
The asymmetry of the transconductance as a function of the gate voltage is obtained in agreement with the experimental data and
the results are explained as due to the depletion of the electron gas within the quantum point contact constriction and the related
reduction of the screening as described within the DFT approach.
The results for transconductance and the $g^*$ factors are in a good agreement with the experimental data [Phys. Rev. B {\bf 81}, 041303, 2010].
\end{abstract}
\maketitle

\section{Introduction}

Quantum point contacts (QPC) -- building blocks for quantum transport
devices -- turned out early to be spin-active elements when
an enhancement of the effective Landé factor $g^{*}$ in QPCs
for the two-dimensional electron gas confined to a quasi one-dimensional constriction was found \cite{Thomas}.
The interaction effects \cite{meirr}, namely the exchange bias \cite{eb1,eb3,eb4,enh,interaqpc,interaqpc2}
are held responsible for this enhancement as well as for the
0.7 conductance anomaly \cite{Thomas,cronenwett}. In systems with
strong spin-orbit coupling the QPCs act as spin-filters working in the absence of the external magnetic field
\cite{Eto,Aharony,Debray,Kim,Nowak,REV,Governale,Datta}.
In the spin control and manipulation devices \cite{Eto,Aharony,Debray,Kim,Nowak,REV,Governale,Datta,chuang}
the external magnetic field can be replaced by an effective momentum-dependent magnetic field \cite{Meier}
introduced by the spin-orbit (SO) interaction for moving carriers.
The SO interaction results from the asymmetry of the structure or
the crystal lattice and in turn introduces a spatial anisotropy of
the spin phenomena, introducing in particular
the conductance dependence on the orientation of the in-plane magnetic
field \cite{Pershin,Scheid,Goulko}.
The anisotropy of the $g^{*}$
factors for an in-plane and an out-of-plane orientation of the magnetic
field is clearly present in the experimental data \cite{Martin,Lu}.
Additionally, the data of Ref. \onlinecite{Martin} exhibited a weak anisotropy
of the $g^{*}$ factor for the in-plane orientation of the magnetic field
i.e. perpendicular and parallel to the direction of the current flow.
A stronger anisotropy of the $g^{*}$ factors for the in-plane orientation of
the magnetic field was observed for holes in quantum wires \cite{chen} and quantum point contacts \cite{nichele}
and explained \cite{magdalena,qpcholes} as due to the spin-orbit interaction
for the valence band carriers. The effective $g^{*}$ factors for electrons in quantum wires was
studied in a number of papers \cite{Debald,gha,kum,sakr}. The combined
effects of SO and electron-electron interactions for 0.7 anomaly
in quantum point contacts were recently discussed \cite{Goulko} for
neglected mixing of subbands within the QPC constriction.

The purpose of the present paper is an analysis of the interactions
and spin-orbit coupling effects for the effective $g^{*}$ factors
in quantum point contacts defined within a two-dimensional electron
gas. While the previous theories dealt with the Zeeman shifts of subbands
\cite{chen,magdalena,qpcholes,Debald,gha,kum,sakr,eb1,eb3,eb4,enh,interaqpc,interaqpc2},
we report on a numerical model focused on simulation of the standard
experimental procedure \cite{Martin,Lu,Martin2,Dena,Patel,Dano} for
extraction of the $g^{*}$ factors from the experimental transconductance.
The procedure \cite{Martin,Lu,Martin2,Dena,Patel,Dano} exploits the transconductance
lines at the edges of half-quantized conductance plateaux that appear in external magnetic
field in presence of the source-drain bias. We analyze the anisotropy
of the $g^*$ factor and its dependence on the subband index for Rashba SO interaction
due to an external electric field.
We find an unexpected reduction of the effective
$g^{*}$ factor for the external magnetic field oriented in-plane and
perpendicular to the current flow direction i.e. parallel to the nominal
orientation of the effective magnetic field due to the SO coupling.
The procedure used for extraction of the $g^{*}$ factors from transconductance
gathers the data of the bottom of subsequent subbands entering the
transport window, for which the effective magnetic field is negligible.
We still find that the extracted $g^{*}$ factors are anisotropic
for the in-plane orientation of the external field, and explain this
feature by the SO coupling of the constriction subbands.

We discuss the $g^*$ factor anisotropy with and without the electron-electron
interaction. For the interaction effects described within the spin
density functional theory (SDFT) approach \cite{eb1,eb3,eb4} we find
a good agreement with the experimental $g^{*}$ factors of Ref. \cite{Martin}.
Moreover, we discuss the interaction effects due to removal of the
electron charge density from the QPC constriction \cite{pinchoff}
and demonstrate that it leads to the asymmetry of the transconductance
lines as a function of the gate voltage as observed experimentally \cite{Martin}. We separate the
contribution of the spin-orbit coupling effects and the orbital effects
in the out-of-plane magnetic field. The electron-electron interaction alone induces a strong imbalance
between the $g^{*}$ factor for the in-plane and out-of plane magnetic
field orientation.

\section{QPC model and transport theory}

\begin{figure}[ht]
\begin{centering}
\includegraphics[width=0.4\paperwidth]{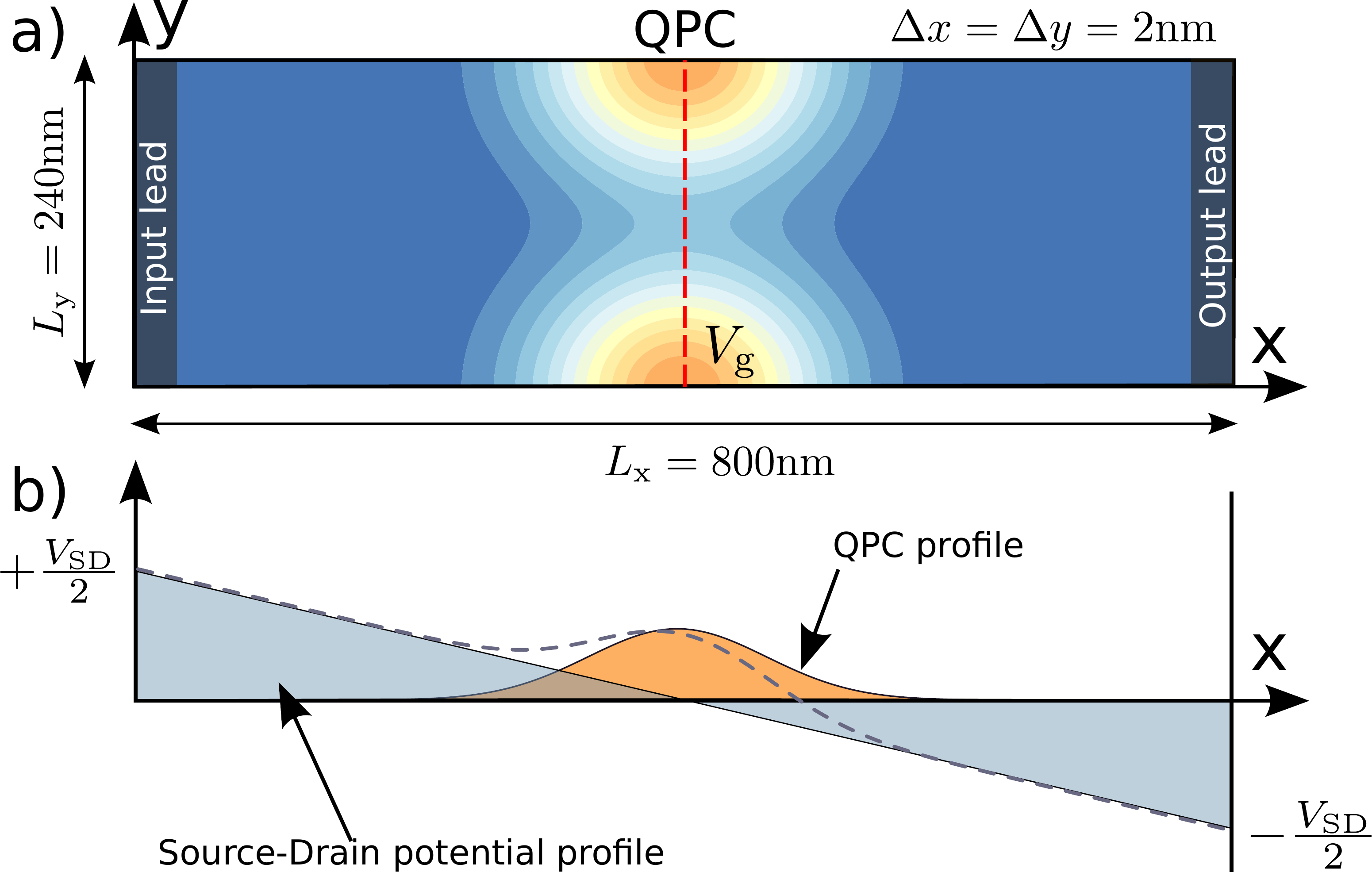}
\par\end{centering}

\caption{\label{fig:uklad}(a) Sketch of the considered system.
A channel of width 240nm with QPC potential modeled  by Eq. (\ref{eq:Vext}).
The channel is assumed infinite in the $x$ direction with the computational
box that covers 800 nm. The red line in the center of the QPC indicates
the cross section of the system for which the dispersion relations
in Fig. \ref{fig:reldysp} were obtained. (b) Cross section of
the system along the $x$-axis showing the potential distribution
for a finite $V_{\mathrm{SD}}$ potential. We assume the symmetric
drop of the potential along the device. The dashed line schematically
shows the total potential profile $V_{\mathrm{SD}}+V_{\mathrm{ext}}$. }
\end{figure}

We consider a system defined within a two-dimensional electron gas
of a wide channel containing a constriction of a quantum point contact
(QPC) in its center (see Fig. \ref{fig:uklad}(a)). The channel's
width is 240nm and its length is assumed infinite (computational box
covers 800 nm). The electrons flow from the input/source lead to the
output/drain lead. We solve the scattering problem of the Fermi level
electrons using the one-electron spin dependent effective band mass
Sch{r}ödinger equation for Fermi energy $E_{F}$
\begin{align}
\left\{ \left[\frac{\boldsymbol{\vec{p}}^{2}}{2m_{\mathrm{eff}}}+eV_{\mathrm{ext}}\right]\boldsymbol{I}+\frac{1}{2}g\mu_{\mathrm{B}}\vec{\boldsymbol{B}}\cdot\boldsymbol{\vec{\sigma}}\right. & +\label{ham}\\
\left.+\boldsymbol{H}_{\mathrm{lat}}+\boldsymbol{H}_{\mathrm{rashba}}\right\} \left(\begin{array}{c}
\chi^{\uparrow}\\
\chi^{\downarrow}
\end{array}\right) & =E_{\mathrm{F}}\left(\begin{array}{c}
\chi^{\uparrow}\\
\chi^{\downarrow}
\end{array}\right),\nonumber
\end{align}
where the plane of confinement is taken at $z=0$ and ${\vec{\boldsymbol{B}}}=(B_{x},B_{y},B_{z})$
is the magnetic field vector. We use the material
parameters for the In$_{0.5}$Ga$_{0.5}$As alloy  \cite{Nowak2014}, the effective mass $m_{\mathrm{eff}}=0.0465m_0$
and the Land\'e factor $g=9$. In Eq. (1) $\boldsymbol{\vec{p}}=\hbar\vec{\boldsymbol{k}}=\hbar\left(-i\boldsymbol{\vec{\nabla}}+\frac{e\boldsymbol{\vec{A}}}{\hbar}\right)$
is the momentum operator and $\boldsymbol{A}$ is the vector potential.
The external potential $V_{\mathrm{ext}}$ describes the QPC narrowing
and is assumed to be of Gaussian shape \cite{Petrovic2015} modeled
with following formula
\begin{align}
V_{\mathrm{ext}} & =V_{\mathrm{gate}}(x,y;400\mathrm{nm},0,50\mathrm{nm},42\mathrm{nm})\nonumber \\
 & +V_{\mathrm{gate}}(x,y;400\mathrm{nm},240\mathrm{nm},50\mathrm{nm},42\mathrm{nm})\label{eq:Vext}
\end{align}
where the upper and lower $V_{\mathrm{gate}}$ terms correspond to
the upper and lower QPC potential profiles (see Fig. \ref{fig:uklad}(a))
and are given by
\[
V_{\mathrm{gate}}(x,y;x_{\mathrm{g}},y_{\mathrm{g}}\Delta_{\mathrm{x}},\Delta_{y})=V_{\mathrm{g}}\mathrm{e}^{-\left(\frac{x-x_{\mathrm{g}}}{2\Delta_{\mathrm{x}}}\right)^{2}}\mathrm{e}^{-\left(\frac{y-y_{\mathrm{g}}}{2\Delta_{\mathrm{y}}}\right)^{2}}.
\]
In the equation above $(x_{\mathrm{g}},y_{\mathrm{g}})$ are the coordinates
of the center of the Gaussian function with $(\Delta_{\mathrm{x}},\Delta_{\mathrm{y}})$
defining its width and height. The QPC constriction is controlled
by the potential $V_{\mathrm{g}}$ parameter which is an analogue
to the electrostatic potential energy introduced by the gate \cite{Martin}.
The third term in the Hamiltonian (\ref{ham}) represents the Zeeman
splitting for magnetic field $\boldsymbol{\vec{B}}$ with $\vec{\boldsymbol{\sigma}}$
being the Pauli matrices vector
\[
\vec{\boldsymbol{\sigma}}=\{\sigma_{\mathrm{x}},\sigma_{\mathrm{y}},\sigma_{\mathrm{z}}\}=\left\{ \left(\begin{array}{cc}
0 & 1\\
1 & 0
\end{array}\right),\left(\begin{array}{cc}
0 & -i\\
i & 0
\end{array}\right),\left(\begin{array}{cc}
1 & 0\\
0 & -1
\end{array}\right)\right\}.
\]
We use the gauge $\boldsymbol{A}=(eB_{y}z-eB_{{z}}y,-eB_{x}z,0)$.
For the plane of confinement $z=0$ the applied vector potential reduces
to the one given by the Landau gauge $\boldsymbol{A}=(-eB_{{z}}y,0,0)$.
Therefore, the in-plane components of the magnetic field $B_{x}$,
$B_{y}$ enter the Zeeman term only. On the other hand $B_{z}$ enters
the momentum operator and introduces orbital effects of the magnetic
field to the kinetic energy and the SO interaction. The latter is described
by the last two terms of the Hamiltonian \cite{REV} with the Rashba
term
\begin{equation}
\boldsymbol{H}_{\mathrm{rashba}}=\gamma_{\mathrm{rsb}}\left\{ \boldsymbol{\sigma}_{\mathrm{x}}\boldsymbol{k}_{\mathrm{y}}-\boldsymbol{\sigma}_{\mathrm{y}}\boldsymbol{k}_{\mathrm{x}}\right\} \label{eq:Hrsb}
\end{equation}
due to the electrostatic confinement of the 2DEG in the growth direction.
The Rashba parameter $\gamma_{\mathrm{rsb}}=\alpha_{\mathrm{3D}}F_{\mathrm{z}}$
defines the strength of the SO interaction in the given material,
where $\alpha_{\mathrm{3D}}$ is a material constant and $F_{\mathrm{z}}$
is the electric field in $z$ direction. We set $\alpha_{\mathrm{3D}}=0.572$nm$^{2}$
as an average value for GaAs and InAs \cite{InGaAsFz1997}.
The values of the electric field in the growth direction are taken up to $F_{\mathrm{z}}=20\frac{\mathrm{meV}}{\mathrm{nm}}$
\cite{InGaAsFz1997}.
The value of $F_{\mathrm{z}}$ can be controlled in the experiment
by the gate voltage. We will discuss below the role of the $\gamma_{\mathrm{rsb}}$
parameter on the anisotropy of the effective g-factor. $\boldsymbol{H}_{\mathrm{lat}}$ in Hamiltonian (1)
describes the so-called lateral SO coupling present due to in-plane
electric fields \cite{REV} with

\[
\boldsymbol{H}_{\mathrm{lat}}=\alpha_{\mathrm{3D}}\boldsymbol{\sigma}_{z}\left\{ \boldsymbol{k}_{\mathrm{x}}\frac{\partial V_{\mathrm{ext}}}{\partial y}-\boldsymbol{k}_{\mathrm{y}}\frac{\partial V_{\mathrm{ext}}}{\partial x}\right\} .
\]
This term is included for consistency, however its influence on the
effective $g^{*}$ factor turns out to be negligible due to
\textit{i)} the symmetry of the considered potential with respect
to the axis of the system \cite{Debray}, \textit{ii)} and small values
of the in-plane gradients $\frac{\partial V_{\mathrm{ext}}}{\partial x_{i}}$ generated
by the QPC potential. 

All the SO effects discussed in this paper result from the electric fields and we
neglect the Dresselhaus SO interaction due to the asymmetry of the crystal lattice.
The neglect of the crystal-asymmetry-induced SO interaction is justified for 
strong vertical electric fields -- assumed below and for large width of the quantum well
confining the electrons in the growth direction \cite{nene}.

In order to solve the scattering problem we use the finite difference
implementation of the quantum transmitting boundary method (QTBM)
\cite{Kirkner1990,Leng1994} for the Sch{r}ödinger equation (\ref{ham}).
A more detailed description of the matrix formulation of the problem
and boundary conditions can be found in the Appendix A. Once the scattering
problem is solved we calculate the conductance using the Landauer
approach. In the linear response regime and for small temperatures
$T\rightarrow0$ the formula can be put in the form
\begin{equation}
G=\frac{e^{2}}{\hbar}T_{\mathrm{tot}}(E_{\mathrm{F}})=\frac{e^{2}}{\hbar}\sum_{k}^{M}T_{k},\label{eq:G}
\end{equation}
where $T_{k}$ is the transmission probability of the $k$-th mode
incoming from the input lead to the system, with $M$ being the total
number of transverse \emph{spinor}-modes. 
 In case of finite source-drain potentials we calculated the current
with the formula
\begin{equation}
I(V_{\mathrm{SD}};T=0)=\frac{e}{h}\intop_{-eV_{\mathrm{SD}}/2}^{+eV_{\mathrm{SD}}/2}T_{\mathrm{tot}}(E_{\mathrm{F}}+E)dE,\label{eq:I}
\end{equation}
where we assume a linear potential drop for non-equal chemical potentials
of the leads. The sketch of the potential profile in the device when
the finite $V_{\mathrm{SD}}$ potential is applied is depicted in
Fig. \ref{fig:uklad}(b). The conductance is calculated by
\begin{equation}
G=\frac{\partial I(V_{\mathrm{SD}})}{\partial V_{\mathrm{SD}}},\label{eq:GI}
\end{equation}
and the transconductance is next defined by $dG/dV_{\mathrm{g}}$.

The finite difference discretization of the Hamiltonian (\ref{ham})
was performed with grid spacing $\Delta x=\Delta y=2$nm
and the Fermi energy $E_{\mathrm{F}}=50$meV. For $\alpha_{\mathrm{3D}}=0$
and $|\vec{\boldsymbol{B}}|=0$ 
the number of transverse modes in the input and output leads at the
Fermi level was $M=$36.

\section{Results without the electron-electron interaction}

\subsection{Dispersion relation}

\begin{figure}[ht]
\begin{centering}
\includegraphics[width=0.38\paperwidth]{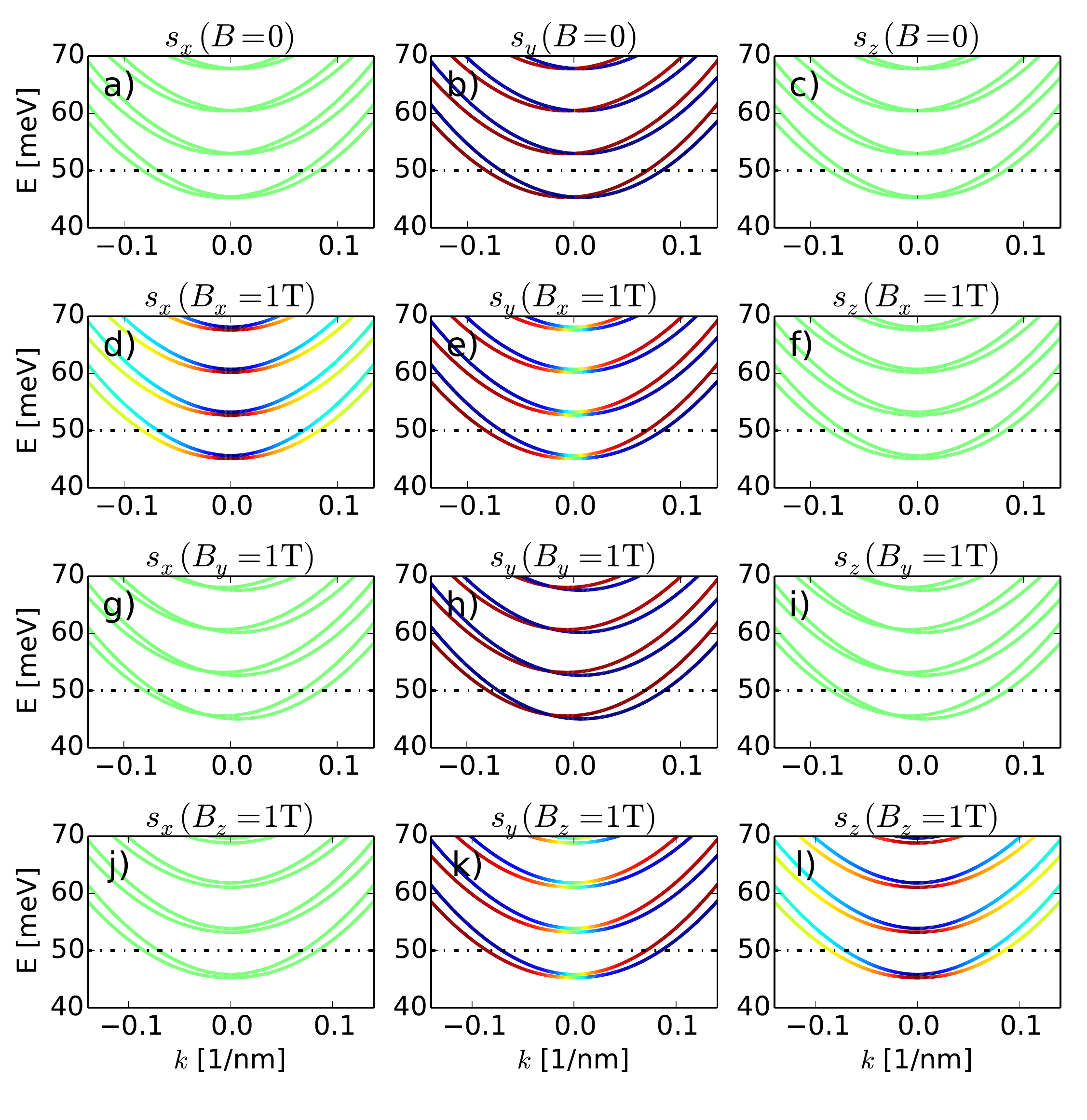}
\par\end{centering}

\caption{\label{fig:reldysp} Dispersion relation for a homogenous channel
of the lateral confinement taken from the QPC constriction (red line
in Fig. \ref{fig:uklad}). Results were obtained for $\alpha_{\mathrm{3D}}=0.572$nm$^{2}$
and $F_{z}=20$ meV/nm which gives $\gamma_{\mathrm{rsb}}=11.4$ meVnm.
The columns indicate the average spin polarization in $x$, $y$ and
$z$ directions (green color stands for $\langle s_{i}\rangle=0$,
the red for $\langle s_{i}\rangle=\hbar/2$, and the blue for $\langle s_{i}\rangle=-\hbar/2$).
The first row (a-c) is for $B=0$. The other plots correspond to $|B|=1.0$
T, oriented in $x$ direction (d-f), $y$ (g-i) and $z$ (j-l).}
\end{figure}

In order to set the ground for the further analysis we display in
Figure \ref{fig:reldysp} the dispersion relation and the spin polarization
of separate subbands for a homogenous channel of lateral confinement
that we adopted from the narrowest cross section of the QPC (red line
in Fig. \ref{fig:uklad}). In the absence of the external magnetic
field (Fig. \ref{fig:reldysp}(a-c)) the electron spins are polarized
by the effective magnetic field resulting from the Rashba interaction.
In a general case the field is of the form $B_{eff}=\gamma_{rsb}(k_{y},-k_{x},0)$
(see Eq. (\ref{eq:Hrsb})). In a channel with potential independent
of $x$, the wave vector $k_{x}=k$ is a well defined quantum number.
On the other hand, $k_{y}$ is an operator that produces a vanishing
contribution in terms of the first-order perturbation, due to a definite
symmetry of the wave functions with respect to the axis of the channel.
Hence, for a symmetric homogenous channel we have $B_{eff}=\gamma_{rsb}(0,-k,0)$.
The dispersion relation of Fig. \ref{fig:reldysp}(b) indicates the
Zeeman-like splitting of subbands with the lower-energy bands spin-polarized
parallel to $B_{eff}$, i.e in the $-y$ direction for $k>0$, and
$+y$ direction for $k<0$. The SO coupling for $B=0$ introduces
shifts of the dispersion relations for $y$-polarized spin subbands
along the wave vector scale \cite{Moroz}, which are next shifted
on the energy scale by the magnetic field oriented in the $y$ direction,
resulting in the asymmetry \cite{Pershin} of the dispersion relation
of Fig. \ref{fig:reldysp}(g-i). The polarization of subbands remains
unchanged when the external magnetic field is applied parallel to
the $y$ axes {[}Fig. \ref{fig:reldysp}(g-i){]}. On the other hand
for the magnetic field parallel to the $x$ and $z$ axis we observe
a competition between the Zeeman effects due to the external field
and the Rashba effective field. The latter prevails at high $k$.

\subsection{Transconductance and the $g^{*}$ factors}

\begin{figure}[ht]
\begin{centering}
\includegraphics[width=0.4\paperwidth]{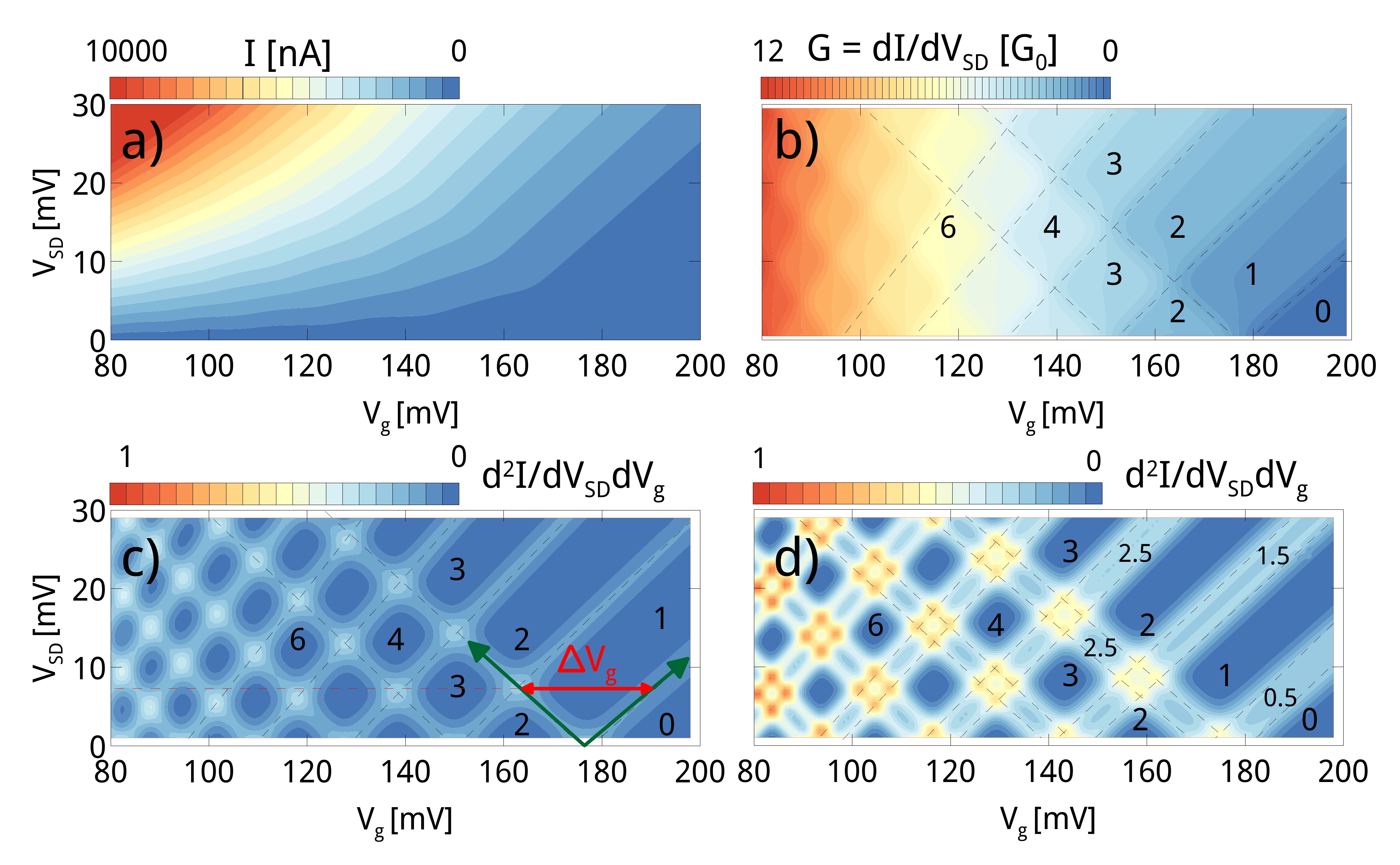}
\par\end{centering}

\caption{\label{fig:vsd} (a) Current as given by the Landauer formula (\ref{eq:I}).
(b) Conductance $G=dI/dV_{sd}$ in units of $G_{0}=\frac{e^{2}}{h}$.
The lines indicate the conductance steps. (c) The transconductance
plot $G=d^{2}I/dV_{sd}dV_{g}$ calculated from (a). The green arrows
indicate the voltage conditions for which the lowest subband follows
the source or drain Fermi levels as the gate voltage $V_{g}$ is varied.
The lines are used to determine the level arm, i.e. the gate-voltage
to energy conversion factor. (a-c) were obtained for $B=0$ and $\gamma_{\mathrm{rsb}}=0$.
(d) The transconductance for $B=2.0$ T.}
\end{figure}

We performed calculations to simulate the classical procedure \cite{Martin,Lu,Martin2,Dena,Patel,Dano}
of extracting the $g^{*}$ factors from the experimental transconductance
$d^{2}I/dV_{\mathrm{g}}dV_{\mathrm{SD}}$ data based on compensation
of the Zeeman splitting by the source-drain bias. Figure \ref{fig:vsd}(a)
shows the current as calculated from the Landauer formula (\ref{eq:I})
for $B=0$. The resulting conductance $G=\frac{dI}{dV_{sd}}$ is displayed
in Figure \ref{fig:vsd}(b) and exhibits characteristic diamond shaped plateaux. The transconductance
of Figure \ref{fig:vsd}(c) is non-zero only near voltages for which
the number of subbands of the constriction that fall between the electrochemical
potentials of source and drain changes. The plateaux of odd conductance
(in units of $G_{0}=\frac{e^{2}}{h}$) appear \cite{Patel} when the
number of conducting subbands for the $+x$ and $-x$ directions differ
by one. A non-zero external magnetic field {[}cf. Fig. \ref{fig:vsd}(d){]}
introduces additionally Zeeman splitting of subbands producing half-plateux
of conductance.

\label{bo} The transconductance as a function of the
voltages is used in the experiments \cite{Martin,Martin2,Patel,Lu}
to extract the gate voltage to energy conversion factor. We use the
data for $B=0$ of Fig. \ref{fig:vsd}(c) to evaluate the conversion
factor
\[
C_{\mathrm{conv}}=\frac{1}{2}\frac{dV_{\mathrm{SD}}}{dV_{\mathrm{g}}}.
\]
The 1/2 factor accounts for the shift of the source (drain) potentials,
which is half of the applied bias: $1/2eV_{sd}$ ($-1/2eV_{sd}$).
The derivative $\frac{dV_{\mathrm{SD}}}{dV_{\mathrm{g}}}$ is evaluated
as the ratio of the source drain voltage to the $\Delta V_{g}$ splitting
of the transconductance lines {for the lowest subband  {[}}see
Fig. \ref{fig:vsd}(c){{]}. The conversion factor depends on the subband index since the subbands differ in their reaction to $V_g$ variation.
The difference results from the specific form of the transverse wave functions for separate subbands.
However, we find that the conversion factors do not depend on the SO coupling constant $\gamma_{rsb}$.
The determined values of the level arm for the first, second and third
subbands $C_{\mathrm{conv}}^{1}={0.27}$, $C_{\mathrm{conv}}^{2}={0.26}$,
$C_{\mathrm{conv}}^{3}={0.32}$ are used below for determination of
the $g^{*}$ factors.

\begin{figure}[H]
\begin{centering}
\includegraphics[width=0.4\paperwidth]{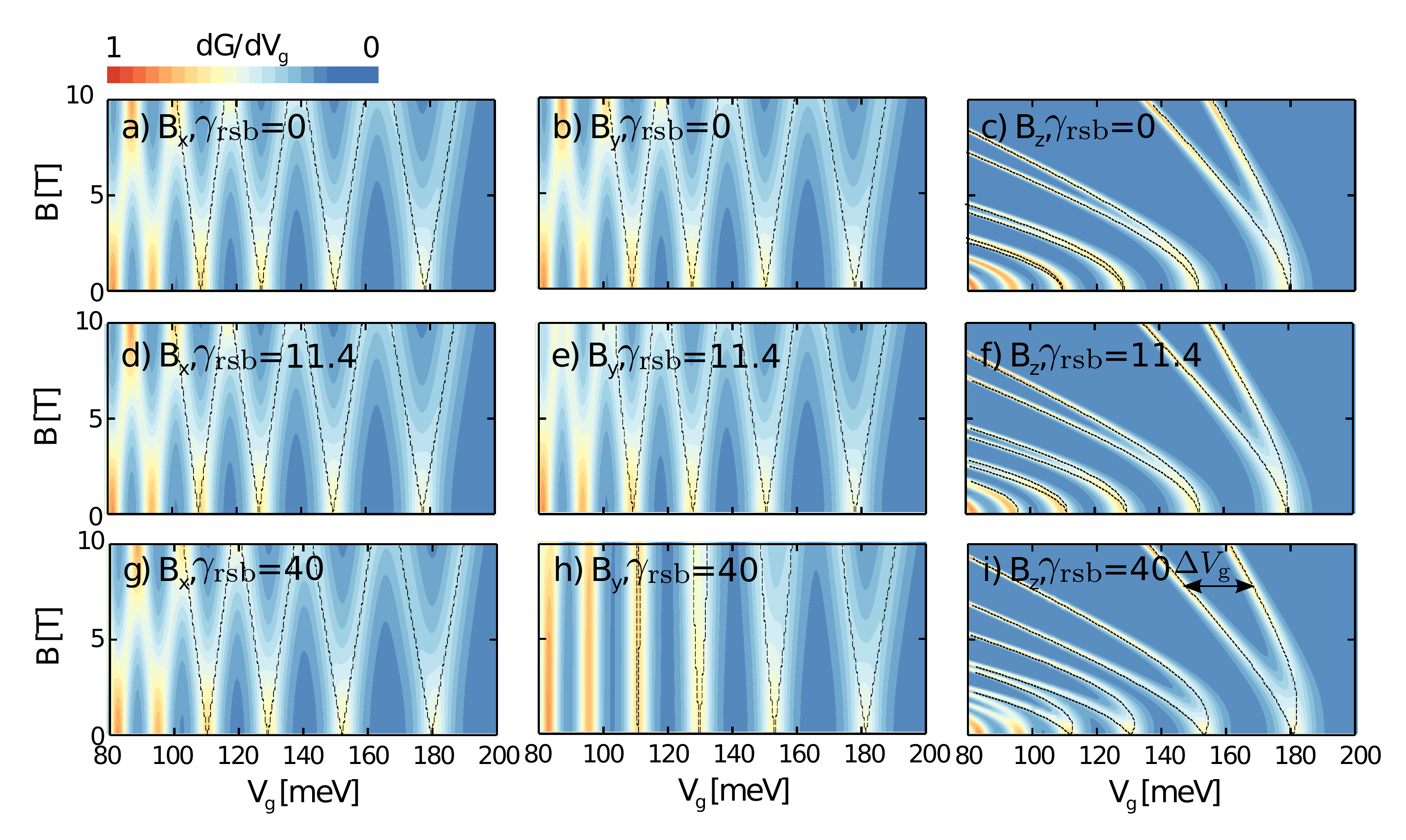}
\par\end{centering}

\caption{\label{fig:vb} Transconductance for $x$ (a,d,g), $y$ (b,e,h) and
$z$ (c,f,i) orientation of the magnetic field, in the absence of
the electron-electron interaction and SO coupling $\gamma_{\mathrm{rsb}}=0$
(a-c), for $\gamma_{\mathrm{rsb}}=11.4$ meVnm (d-f) and $\gamma_{\mathrm{rsb}}=40$
meVnm (g-i). The dashed lines indicate conditions for which one obtains
$k_{x}=0$ at the Fermi energy for a homogenous infinite channel with
the lateral confinement adopted from the center of the constriction.}
\end{figure}

As a next step in the procedure, the transconductance is evaluated
as a function of the magnetic field. The results with and without
the SO coupling are displayed in Fig. \ref{fig:vb}. We read out the
Zeeman splitting of the transconductance lines in external magnetic
field on the gate voltage $\Delta V_{g}$ scale. The $g_{k}^{*}$
factor for $k$-th subband is then evaluated as
\begin{equation}
g_{k}^{*}=\frac{1}{\mu_{B}}\frac{d(\Delta V_{\mathrm{g}}(B))}{dB}C_{\mathrm{conv}}^{k}.\label{eq:gf}
\end{equation}
The splitting of transconductance lines is plotted in Fig. \ref{fig:dvb},
and the calculated slope determines the effective Landé factors.

For $B=(B_{x},0,0)$, the SO coupling leaves the splitting of the
transconductance lines $\Delta V_{g}$ unchanged {[}Fig. \ref{fig:vb}(a,d,g)
and Fig.\ref{fig:dvb}(a,d){]}. On the other hand the splitting $\Delta V_{g}$
for the magnetic field oriented in-plane but perpendicular to the
current flow $B=(0,B_{y},0)$ is decreased by the SO coupling {[}Fig.
\ref{fig:vb}(b,e,h) and Fig. \ref{fig:dvb}(b,e){]}. The effective
magnetic field for electrons moving in the $x$ direction polarizes
their spins in the $y$ direction. Therefore one should expect that
SO interaction will leave the Zeeman effect for $y$ orientation of
the field unchanged. Moreover, since for $B=0$ the spins of moving
electrons are already parallel or antiparallel to $y$ axis {[}see
Fig. \ref{fig:reldysp}{]}, $\Delta V_{g}(B_{x})$ should be expected
decreased by SO interaction. Nevertheless the calculated results are
just opposite. The transconductance lines for $B_{x}$ tends to cross
at higher magnetic field. No anti-crossing is observed for this orientation
of the magnetic field, as should be expected for the competition between
spin polarization between $B_{x}$ and the SO effective magnetic field.

Summarizing, the results of Figure \ref{fig:gf}: In the absence of
the spin-orbit interaction the evaluated in-plane $g^{*}$ factor
is isotropic for the in-plane orientation of the field and is nearly
equal to the material constant applied in the Hamiltonian $g=9$, and
subband independent. The orbital effects of the perpendicular magnetic
field $B_{z}$ due to the potential introduced by the field within
the constriction enhance the evaluated $g^{*}$ factors for higher
subbands also when SO coupling is absent. For the conversion factor
which increases with the subband index, the orbital effects keep the
$\Delta V_{g}$ unchanged for subsequent subbands in Fig. \ref{fig:dvb}(c).
Fig. \ref{fig:gf} indicates, that the \textit{(i)} spin-orbit interaction
enhances the $g_{z}^{*}$ factor \textit{(ii)} introduces in-plane
asymmetry of the $g_{x}^{*},g_{y}^{*}$ factor \textit{(iii)} leaves
unchanged $g_{x}^{*}$ and \textit{(iv)} reduces $g_{y}^{*}$. We
analyze these effects in the next Section.

\begin{figure}[H]
\begin{centering}
\includegraphics[width=0.4\paperwidth]{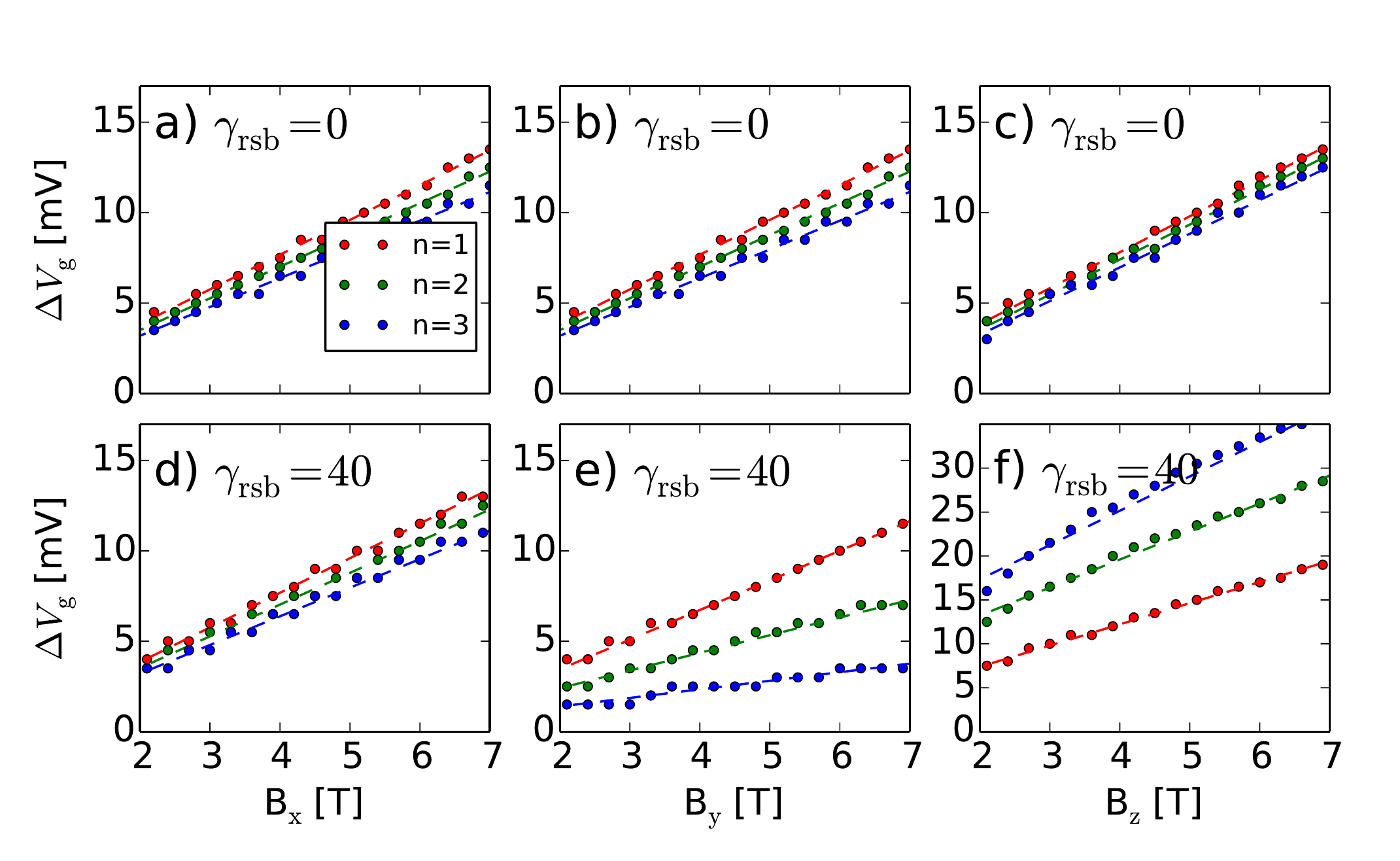}
\par\end{centering}

\caption{\label{fig:dvb} The splitting of the transconductance lines as read
out of Fig. \ref{fig:vb} for the magnetic field oriented in the $x$
(a,d), $y$ (b,e), and $z$ (c,f) directions. The electron-electron
interaction is neglected in these data.}
\end{figure}

\begin{figure}[H]
\begin{centering}
\includegraphics[width=0.4\paperwidth]{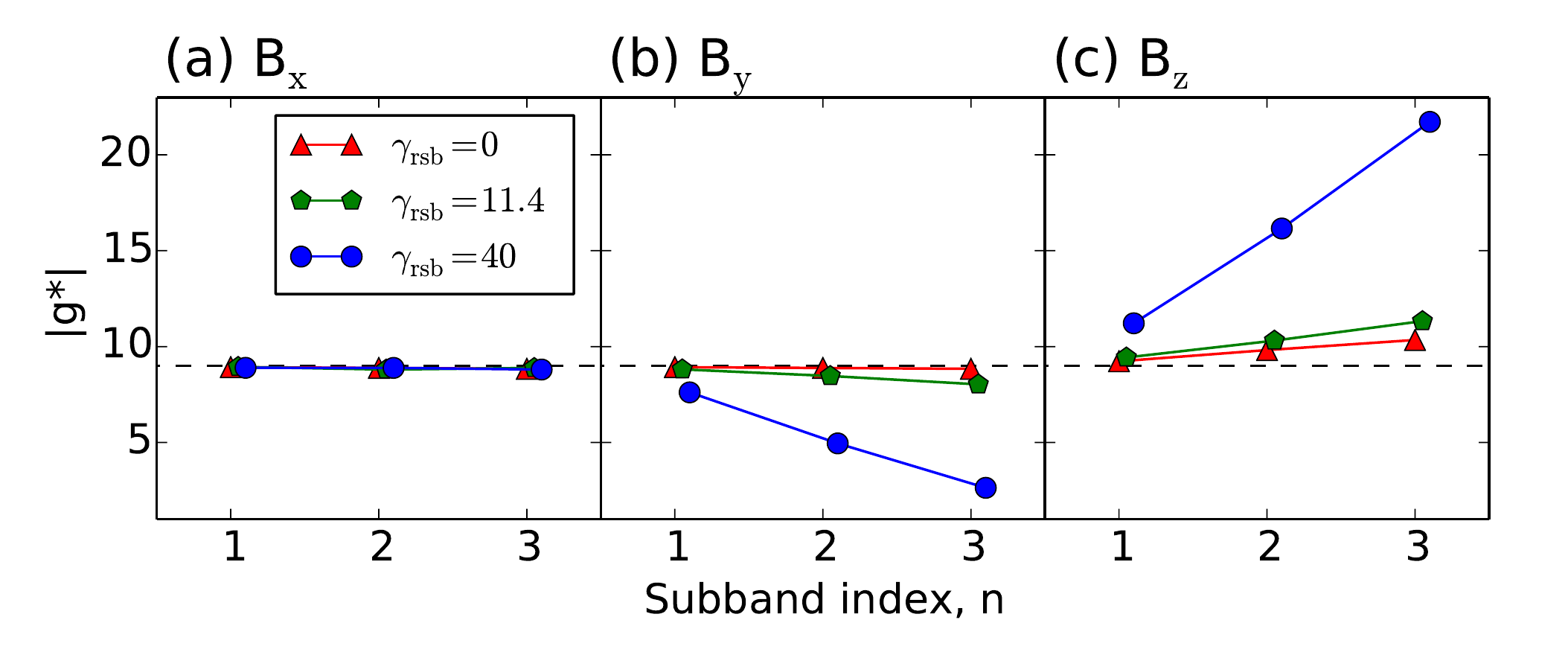}
\par\end{centering}

\caption{\label{fig:gf} The calculated $g^{*}$ factors for various orientations
of the magnetic field and spin-orbit coupling constant. The electron-electron
interaction is neglected in the plotted results.}
\end{figure}

\section{SO effect on observed $g^{*}$ factors}

\subsection{Neglected subband mixing}

\label{secx} In order to explain the results presented above, let
us consider a homogenous channel with the lateral confinement given
by a harmonic oscillator potential
\begin{equation}
U_{ho}=\frac{1}{2}m_{\mathrm{eff}}\omega^{2}y^{2}.\label{eq:Uho}
\end{equation}
The spatial electron wave functions in the absence of SO interaction
for the $n$-th subband is of the form
\begin{equation}
\psi_{n}(x,y)=A_{n}e^{-\frac{m_{\mathrm{eff}}\omega y^{2}}{2\hbar}}H_{n}\left(\sqrt{\frac{m_{\mathrm{eff}}\omega}{\hbar}}y\right)\, e^{-ik_{x}^{n}\, x},\label{eq:rzbaza}
\end{equation}
with normalization constant $A_{n}=\frac{1}{\sqrt{2^{n}n!}}\left(\frac{m_{\mathrm{eff}}\omega}{\pi\hbar}\right)^{\frac{1}{4}}$
and $H_{n}$ being the n-th Hermite polynomial. The transconductance
that is used in the procedure for extracting the $g^{*}$ factors
is non-zero when one of the subbands enters or leaves the transport
window defined by the source and drain potentials, hence for further
analysis we take $k_{x}^{n}=0$ (this condition follows the transconductance
steps as evaluated from derivatives of the current - see Fig. \ref{fig:vb}.)

We consider the effects of SO interaction for weak magnetic fields.
We neglect the term that is quadratic in $B_{z}$ for the magnetic
field oriented perpendicular to the plane of confinement. We adopt
the spin basis of $\sigma_{z}$ operator. The position of the bottom
of subsequent subbands for $k_{x}=0$ is $E_{n}=\hbar\omega\left(n+\frac{1}{2}\right)$.
For a narrow channel when the value of $\hbar\omega$ is high the energy
spacings between the subbands are large. Therefore, it seems justified
to neglect the subband mixing \cite{Goulko}. We diagonalize the SO
Hamiltonian in the basis of the spin-up and spin-down basis of the
lowest subband ($n=1$),
\begin{align}
\phi_{0,\uparrow} & \equiv A_{0}e^{-\frac{m_{\mathrm{eff}}\omega y^{2}}{2\hbar}}H_{0}\left(\sqrt{\frac{m_{\mathrm{eff}}\omega}{\hbar}}y\right)\left(\begin{array}{c}
1\\
0
\end{array}\right),\label{eq:b11}\\
\phi_{0,\downarrow} & \equiv A_{0}e^{-\frac{m_{\mathrm{eff}}\omega y^{2}}{2\hbar}}H_{0}\left(\sqrt{\frac{m_{\mathrm{eff}}\omega}{\hbar}}y\right)\left(\begin{array}{c}
0\\
1
\end{array}\right).\label{eq:b12}
\end{align}
For analysis we consider the Hamiltonian $\boldsymbol{H}$ (\ref{ham})
(with $U_{ho}$ term (\ref{eq:Uho}) replacing $V_{ext}$), in which the lateral
SO field is missing due to the vanishing of the term $\frac{\partial V_{\mathrm{ext}}}{\partial x}$
and $k_{x}=0$. For $k_{x}=0$, the Rashba SO interaction reduces
to $\boldsymbol{H}_{\mathrm{rashba}}=\gamma_{\mathrm{rsb}}\left\{ \boldsymbol{\sigma}_{x}\boldsymbol{k}_{y}-\boldsymbol{\sigma}_{y}\left(-eB_{\mathrm{z}}y\right)\right\} $.
The Hamiltonian matrix for the above basis
\begin{equation}
H_{\sigma'\sigma}=\left\langle \phi_{\sigma'}\left|\boldsymbol{H}\right|\phi_{\sigma}\right\rangle ,\label{eq:Hij}
\end{equation}
takes the form
\begin{equation}
\boldsymbol{H}_{n}\equiv\left[\begin{array}{cc}
E_{n}+\frac{g\mu_{\mathrm{B}}B_{\mathrm{z}}}{2} & \frac{g\mu_{\mathrm{B}}\left(iB_{\mathrm{y}}+B_{\mathrm{x}}\right)}{2}\\
\frac{g\mu_{\mathrm{B}}\left(B_{\mathrm{x}}\text{\textminus}iB_{\mathrm{y}}\right)}{2} & E_{n}-\frac{g\mu_{\mathrm{B}}B_{\mathrm{z}}}{2}
\end{array}\right].\label{eq:Hn}
\end{equation}
From the form of the Hamiltonian it is evident, that for neglected subband mixing the effect of the Rashba
interaction does not affect the Hamiltonian matrix. Thus the eigenvalues
of the Hamiltonian above are same for three perpendicular orientations
of the magnetic field:
\begin{align*}
E_{\sigma}^{\mathrm{x}}\left((B,0,0)\right) & =E_{\sigma}^{\mathrm{y}}\left((0,B,0)\right)=E_{\sigma}^{\mathrm{z}}\left((0,0,B)\right)\\
 & =E_{\mathrm{0}}+\sigma\frac{g\mu_{\mathrm{B}}B}{2}.
\end{align*}

\subsection{Subband mixing effect}

In order to describe the SO effects on the energies of the bottom
of the subbands for the in-plane magnetic field orientation one needs
to account for the subband mixing. We add the second subband to the
basis
\begin{align}
\phi_{1,\uparrow} & \equiv A_{1}e^{-\frac{m_{\mathrm{eff}}\omega y^{2}}{2\hbar}}H_{1}\left(\sqrt{\frac{m_{\mathrm{eff}}\omega}{\hbar}}y\right)\left(\begin{array}{c}
1\\
0
\end{array}\right),\label{eq:b23}\\
\phi_{1,\uparrow} & \equiv A_{1}e^{-\frac{m_{\mathrm{eff}}\omega y^{2}}{2\hbar}}H_{1}\left(\sqrt{\frac{m_{\mathrm{eff}}\omega}{\hbar}}y\right)\left(\begin{array}{c}
0\\
1
\end{array}\right).\label{eq:b24}
\end{align}
The Hamiltonian matrix (\ref{eq:Hij}) in the basis of four states
with $n=0$ and $n=1$ {[}Eq. (\ref{eq:b11}), (\ref{eq:b12}), (\ref{eq:b23}),
and (\ref{eq:b24}){]} takes the form
\[
\boldsymbol{H}\equiv\left[\begin{array}{cc}
\boldsymbol{H}_{0} & \boldsymbol{H}_{\mathrm{01}}\\
\boldsymbol{H}_{\mathrm{01}}^{\dagger} & \boldsymbol{H}_{1}
\end{array}\right],
\]
where $\boldsymbol{H}_{\mathrm{0}}$ and $\boldsymbol{H}_{\mathrm{1}}$
are given by (\ref{eq:Hn}) and {\footnotesize{
\[
\boldsymbol{H}_{01}\equiv\left[\begin{array}{cc}
0 & \frac{-i\gamma_{\mathrm{rsb}}}{\sqrt{2\hbar m_{\mathrm{eff}}\omega}}\left(m_{\mathrm{eff}}\omega-e\hbar B_{\mathrm{z}}\right)\\
\frac{-i\gamma_{\mathrm{rsb}}}{\sqrt{2\hbar m_{\mathrm{eff}}\omega}}\left(m_{\mathrm{eff}}\omega+e\hbar B_{\mathrm{z}}\right) & 0
\end{array}\right].
\]
}}Note, that for $B_{z}=0$ the SO constant $\gamma_{\mathrm{rsb}}$
gets effectively enhanced for a larger values of $\hbar\omega$. Now, for
$B=(B_{x},0,0)$ the eigenvalues are
\[
E_{\mathrm{n},\sigma}^{\mathrm{x}}=\hbar\omega-\frac{(-1)^{n}}{2}\sqrt{\frac{2m_{\mathrm{eff}}\omega\gamma_{\mathrm{rsb}}^{2}}{\hbar}+\left(\hbar\omega\right)^{2}}+\sigma\frac{g\mu_{\mathrm{B}}B_{\mathrm{x}}}{2}.
\]
The Zeeman splitting for a given $n$ is still not affected by the
SO interaction -- in agreement with Fig. \ref{fig:gf}. For $B=(0,B_{y},0)$
one finds
\begin{align}
E_{\mathrm{n},\sigma}^{\mathrm{y}}=\hbar\omega-\frac{(-1)^{n}}{2}\sqrt{\frac{2m_{\mathrm{eff}}\omega\gamma_{\mathrm{rsb}}^{2}}{\hbar}+\left(\hbar\omega+\sigma g\mu_{\mathrm{B}}B_{\mathrm{y}}\right)^{2}}.\label{eq:Ey}
\end{align}
For nonzero $\gamma_{\mathrm{rsb}}$ the spin splitting is reduced
{[}see Fig. \ref{ww}{]}, in agreement with the results of Fig. \ref{fig:gf}.
For $B=(0,0,B_{z})$ one obtains
\begin{align}
E_{\mathrm{n},\sigma}^{\mathrm{z}} & =\hbar\omega-\frac{(-1)^{n}}{2}\left[\left(\hbar\omega+\sigma g\mu_{\mathrm{B}}B_{\mathrm{z}}\right)^{2}+\right.\nonumber \\
 & \left.\frac{2m_{\mathrm{eff}}\gamma_{\mathrm{rsb}}^{2}}{\hbar^{3}\omega}\left(\hbar\omega+\frac{\sigma e\hbar^{2}B_{\mathrm{z}}}{m_{\mathrm{eff}}}\right)^{2}\right]^{\frac{1}{2}}.\label{eq:Ez}
\end{align}

For the $B_{\mathrm{z}}$magnetic field the SO interaction is more
pronounced when the confinement strength is smaller ($\hbar\omega\rightarrow0$).

\begin{figure}[H]
\begin{centering}
\includegraphics[width=0.3\paperwidth]{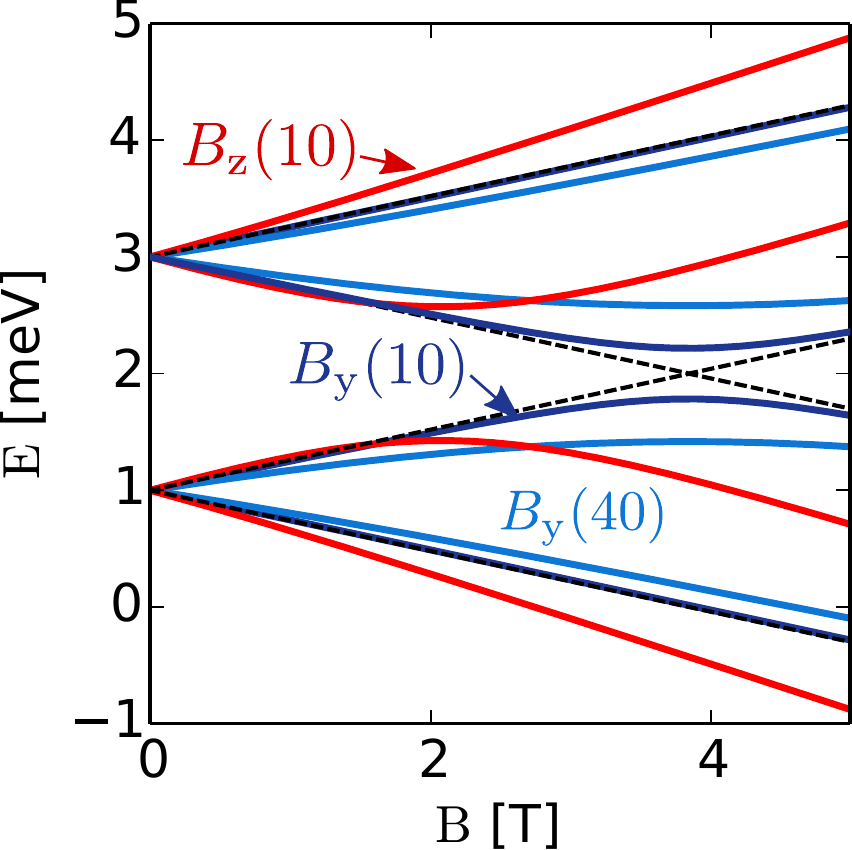}
\par\end{centering}

\caption{\label{fig:dftg-1} \label{ww}Energy levels as calculated in the
basis of Eq. (\ref{eq:b11}), (\ref{eq:b12}), (\ref{eq:b23}), and
(\ref{eq:b24}) for $\hbar\omega=2$meV. Black dashed lines indicate the
results obtained in the absence of the SO coupling. The results are
identical to the ones obtained for $B=(B_{x},0,0)$ in the presence
of the SO coupling -- with a precision of a $B$-independent energy
shift of the energy levels. The red lines show the results for $B=(0,0,B_{z})$
and $\gamma_{rsb}=\alpha_{3D}F_{z}=10$ meVnm. The light and dark
blue lines indicate the results for $B=(0,B_{y},0)$ for $\gamma_{rsb}=10$
meVnm and $\gamma_{\mathrm{rsb}}=40$ meVnm, respectively. A constant
shift of the energy levels for different values of $\gamma_{rsb}$
was applied.}
\end{figure}

The numerical data are given in Fig. \ref{ww} for $\hbar\omega=2$meV.
The calculation reproduces all the features observed in Fig. \ref{fig:gf}
listed at the end of the previous subsection.

We conclude that, the experimental procedure for extraction of $g^{*}$
factors from transconductance detects the bottom of the subbands for
which $k_{x}=0$. For $k_{x}=0$ and $B_{y}=B_{z}=0$, $\sigma_{x}$
commutes with the Rashba and total Hamiltonian. Therefore, for $B=(B_{x},0,0)$
the Hamiltonian eigenstates remain eigenstates of $\sigma_{x}$ operator
even in the presence of the SO interaction. Hence, for the magnetic
field oriented in the $B_{x}$ direction the SO coupling does not
influence the spin states nor the $g^{*}$ factor. The crossing of
transconductance lines of Fig. \ref{fig:vb} for $B_{x}$ follows
(see Fig. \ref{fig:vb}(g)). Hence, the spins at the bottom of the
subbands get oriented by the SO interaction in the $\pm x$ direction
and not in $\pm y$ direction as expected from the orientation of
the effective SO magnetic field. For $B_{y}\neq0$ the splitting of
energy levels is reduced by the SO interaction. Moreover, the description
of the SO effects at the bottom of the subbands and the resulting
anisotropy of $g^{*}$ factor requires account taken for the subband
mixing. The approximation of a single subband \cite{Goulko} is not
suitable for evaluation of the $g^{*}$ factors modification by the SO interaction
as calculated by the main simulation of this paper, and 
as observed in the
transport experiments \cite{Martin,Martin2,Patel,Lu}. 
Moreover, for
$E_{F}$ between $E_{1}$ and $E_{2}$ the SO coupling effects involve
coupling of a transmitting $(n=1)$ and evanescent $(n=2)$ modes,
i.e. the subband mixing is present also when one of the subbands is
above the Fermi level.

\section{Interaction effects}

For evaluation of the interactions effects on the $g^*$ factors we used the
density function theory \cite{Stopa1996}. The calculations for the
interacting systems were performed for somewhat smaller computational
box of length $L_{\mathrm{x}}=600$nm and width $L_{\mathrm{y}}=160$nm
with grid spacing $\Delta x=4$nm, and $N=150$ electrons inside the
2DEG and the QPC confining potential given by a similar formula as
in Eq. (\ref{eq:Vext})
\begin{align*}
V_{\mathrm{ext}} & =V_{\mathrm{gate}}(x,y;300\mathrm{nm},0,50\mathrm{nm},34\mathrm{nm})\\
 & +V_{\mathrm{gate}}(x,y;300\mathrm{nm},160\mathrm{nm},50\mathrm{nm},34\mathrm{nm}).
\end{align*}
The approach used in this paper is a development of the one used in
Ref. \cite{szafranDFT2011}, with an extension to spin-orbit coupling
effects. The details are given in Appendix B. The DFT is used to evaluate
the Fermi energy, and the spin-dependent potential landscape. The
calculation is performed as a function of $V_{g}$ and $B$ for $V_{sd}\simeq0$.
For larger $V_{sd}$ we introduce the bias as an additional linear
term as in Section \ref{bo}.

The transconductance plot with the electron-electron interaction included is given in Fig. \ref{fig:dftvsd} for $B=0$.
 The electron-electron interaction introduces asymmetry in the transconductance lines as
functions of the gate-voltage, clearly visible for the first conductance
step. A different path of the lines for $V_{g}$ below or above the
QPC conductance step is related to the appearance or disappearance of
the electron density related to a specific subband depending on the
gate voltage {[} Fig. \ref{kkj}(a){]}. For $V_{g}>250$ meV the electron
density is removed from the center of the QPC {[}Fig. \ref{kkj}(c){]}
and the screening of the gate potential by the electron gas is missing,
hence a larger slope of potential in Fig. \ref{kkj}(b). Similar effects
of the pinch-off for the screening were obtained in Ref.
\cite{pinchoff}, in which the transconductance was not disussed.
The present paper indicates their consequences for
asymmetry of the transconductance lines. The asymmetry of the transconductance
is the strongest for the first subband {[}Fig. \ref{fig:dftvsd}{]},
but appears also for higher subbands, which is accompanied by deviation
of the potential landscape from a linear dependence on $V_{g}$ {[}Fig.
\ref{kkj}(b){]}. In the experiment \cite{Martin} the transconductance
lines are \textit{(i)} asymmetric with $V_{g}$, \textit{(ii)} get
steeper at the higher absolute value of the gate voltage side, and
\textit{(iii)} the asymmetry is reduced for larger $n$. Our calculation
reproduces all these features and explains that their origin is the
screening of the gate potential by the two-dimensional electron gas.
The evaluated conversion factors are $C_{\mathrm{conv}}^{k}=\{0.14,0.10,0.12\}$
for the first, second, and third subbands, respectively. Note, that
the reduction of the conversion factors and their variation with the
subband index is the effect of the gate potential screening.

\begin{figure}[ht]
\begin{centering}
\includegraphics[width=0.4\paperwidth]{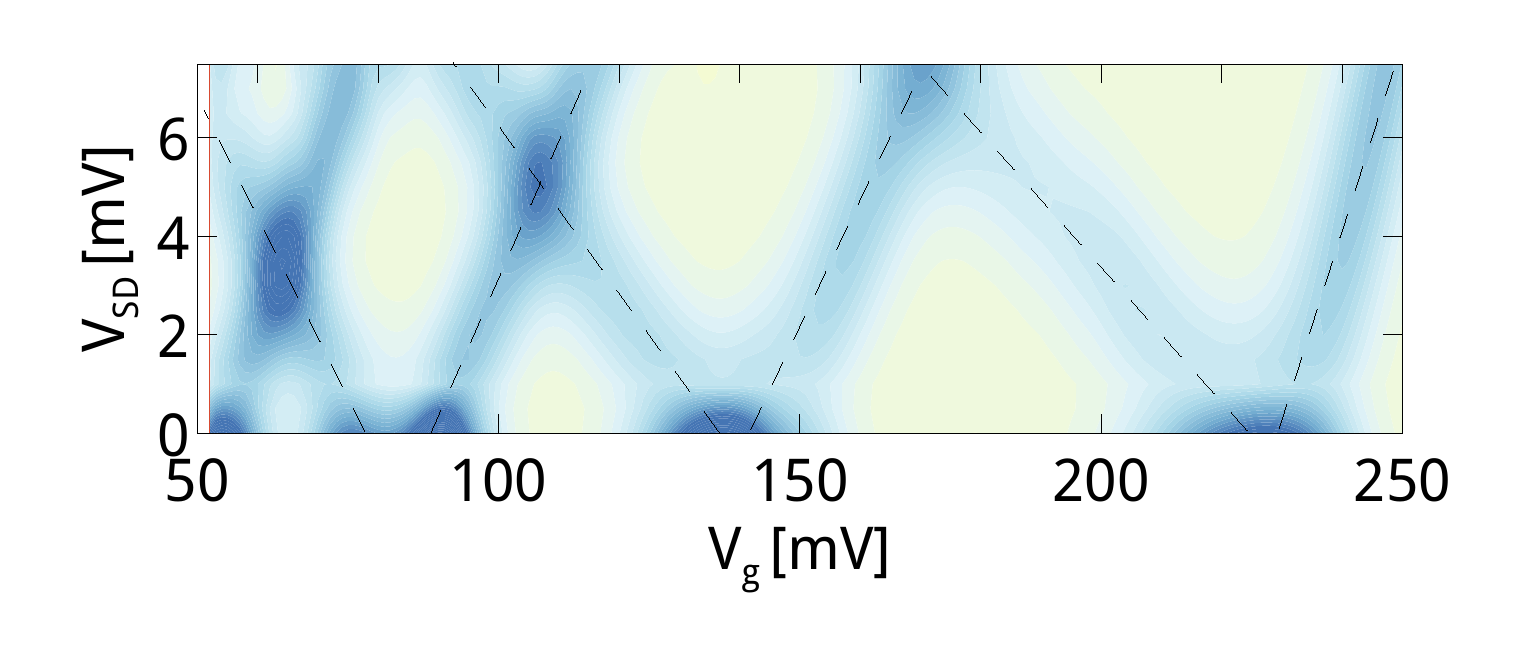}
\par\end{centering}

\caption{\label{fig:dftvsd} Transconductance for $\gamma_{\mathrm{rsb}}=0$
as obtained with the electron-electron interactions described by DFT.}
\end{figure}

\begin{figure}[ht]
\begin{centering}
\includegraphics[width=0.4\paperwidth]{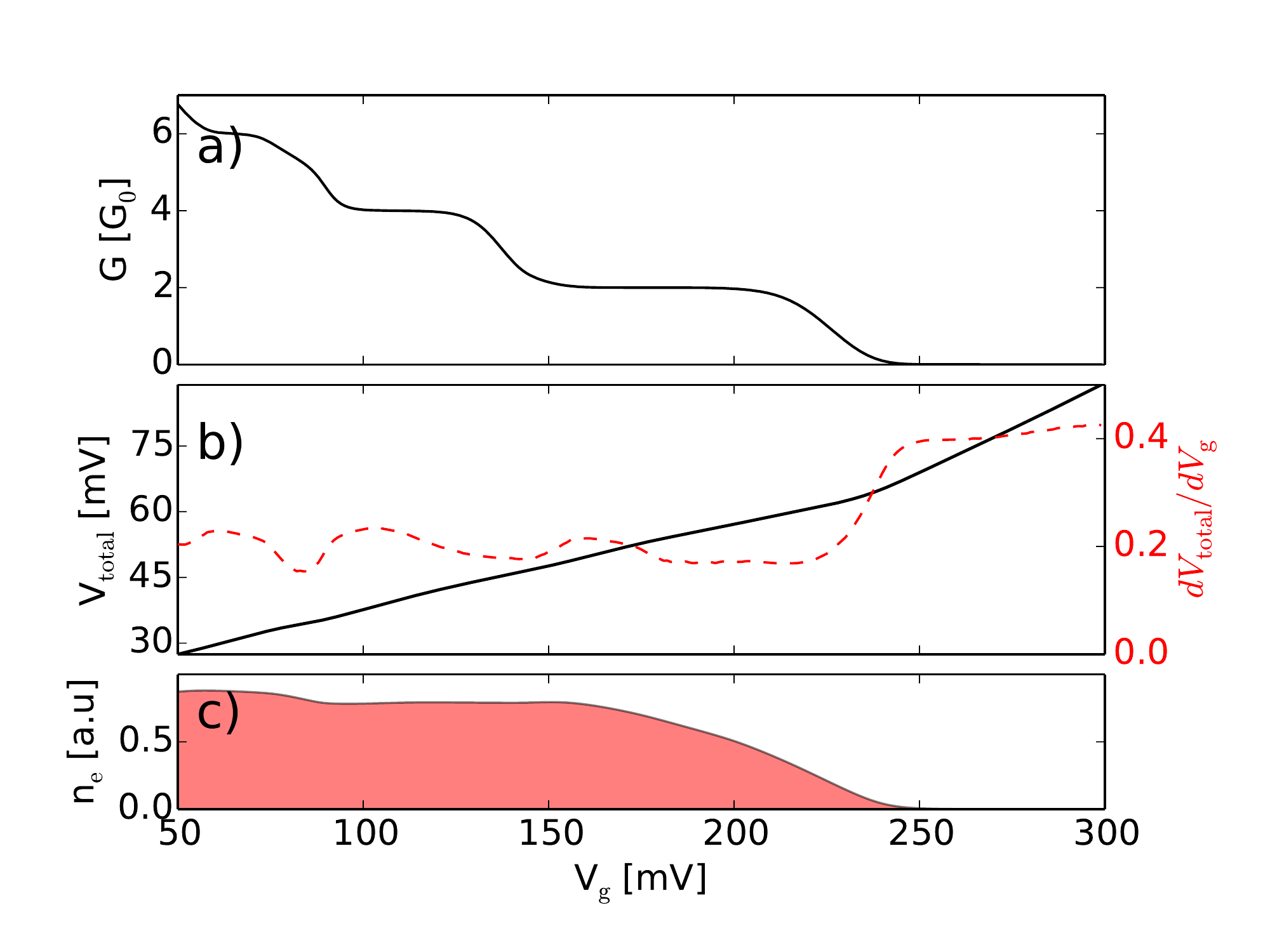}
\par\end{centering}

\caption{(a) The conductance as a function of the gate voltage for $V_{sd}=0$
corresponding to $\ref{fig:dftvsd}$(a). The effective potential obtained
from self-consistent DFT calculations (b) and the electron density
(c) at the center of QPC constriction.}

\label{kkj}
\end{figure}

The transconductance dependence on the magnetic field results is given
in Fig. \ref{fig:dftvgb}, with the positions of the maxima marked
by blue shadows. The evaluated effective $g^{*}$ factors are displayed
in Fig. 11. The
features introduced by the electron-electron interaction are revealed
by comparison with Fig. 6 of Sec. \ref{bo}. We find \textit{(i)} An enhancement
of the $g^{*}$ above the nominal value $g=9$ for the first $n=1$
subband index for each direction of the magnetic field. \textit{(ii)}
The largest enhancement (a factor of $3$) is found for perpendicular
magnetic field in case $B=(0,0,B_{\mathrm{z}})$. In the absence of
SO interaction the in-plane Lande factors $g_{\mathrm{x}}^{*}$ and
$g_{\mathrm{y}}^{*}$, exhibit the same behavior as a function of
$n$. When the Rashba SO in introduced with $\gamma_{\mathrm{rsb}}=11.4\,\mathrm{meVnm}$
the value of the $g_{\mathrm{y}}^{*}$ drops faster with $n$ than
$g_{\mathrm{x}}^{*}$ leading to an anisotropy for in-plane magnetic
field.
The results, including the last feature remain in
a good agreement with the experiment of Ref. \cite{Martin}.
The $g_{\mathrm{x}}^{*}$ / $g_{\mathrm{y}}^{*}$  anisotropy is weak in Ref. \cite{Martin}. A stronger
anisotropy should be expected to appear in InSb quantum point contacts \cite{insb}.

For larger value of $\gamma_{\mathrm{rsb}}=40\,\mathrm{meVnm}$ \textit{(i)}
the effective $g_{\mathrm{x}}^{*}$ is almost the same as for other
Rashba parameters, with a small but visible shift towards larger values.
\textit{(ii)} $g_{\mathrm{y}}^{*}$ drops in a more distinct manner with $n$.
This in an effect encountered already in the previous subsection for the model without
the electron-electron interaction. For the out-of-plane magnetic field
we observe a strong enhancement of the $g_{z}^{*}$ factor with respect
to the in-plane $g_{x}^{*}$ and $g_{y}^{*}$ factors in agreement
with Ref. \cite{Martin}. We obtain more or less a constant behavior
in $n$ for the $\gamma_{\mathrm{rsb}}=0$. For larger SO coupling
constant the $g_{{z}}^{*}$ tends to grow with $n$. In the experiment
\cite{Martin} the $g_{\mathrm{z}}^{*}$ drops with $n$. In Fig.
\ref{fig:gf}(c) we have shown that even without electron-electron interaction the orbital
effects may lead to increment of $g^{*}$ for higher subbands. Additionally,
those effects depend highly on the geometry of the sample thus, one
may expect, that the effect will differ from sample to sample. On
the other hand the results for $g_{{x}}^{*}$, $g_{{y}}^{*}$ are
free of orbital effects and the results should not be sample dependent,
hence good agreement of the present results to the experimental data.

\begin{figure}[H]
\begin{centering}
\includegraphics[width=0.4\paperwidth]{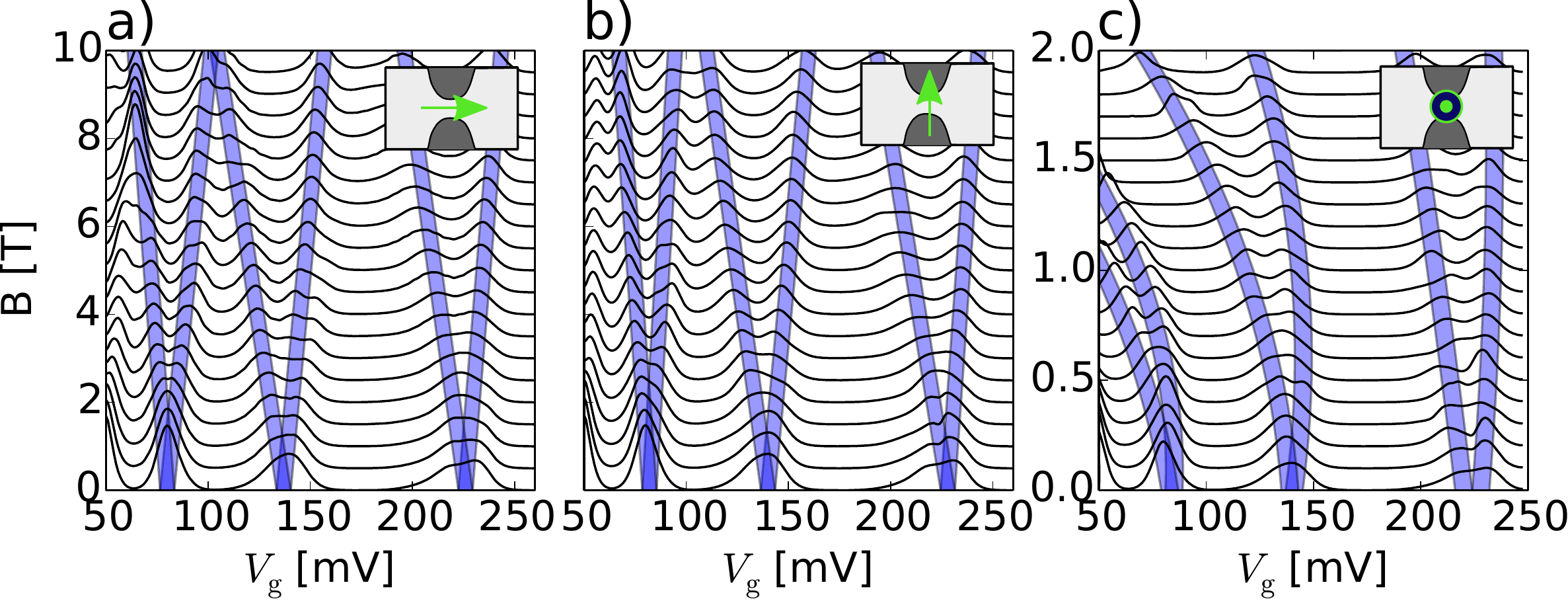}
\par\end{centering}

\caption{\label{fig:dftvgb}The transconductance as a function of magnetic
field, for (a) $B=(B_{\mathrm{x}},0,0)$ (b) $B=(0,B_{\mathrm{y}},0)$
and (c) $B=(0,0,B_{\mathrm{z}})$ for $\gamma_{\mathrm{rsb}}=11.4\,\mathrm{nm}^{2}$.
The insets show schematically the orientation of magnetic field for
each case. The blue lines indicate the positions of the transconductance
peaks used for the evaluation of the $g^{*}$. }
\end{figure}

\begin{figure}[H]
\begin{centering}
\includegraphics[width=0.4\paperwidth]{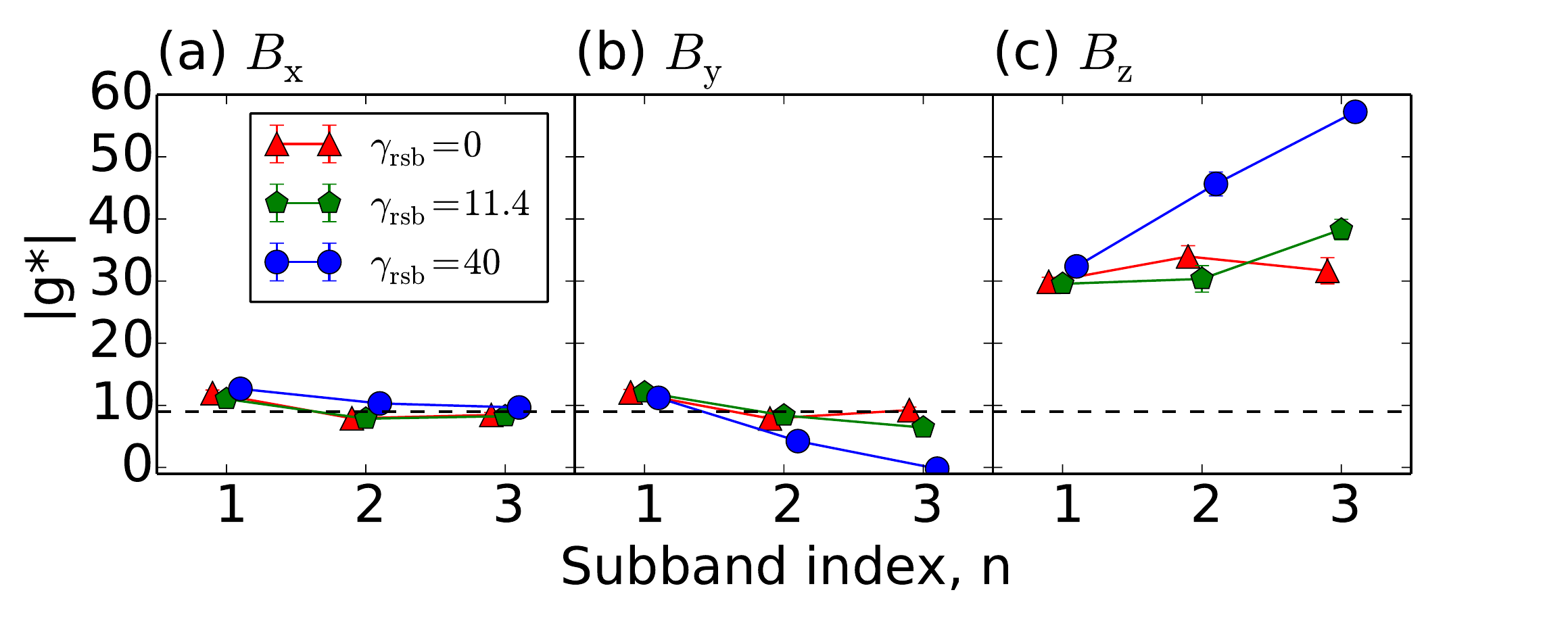}
\par\end{centering}

\caption{\label{fig:dftg} The calculated values of $|g^{*}|$ for three different
orientations of the magnetic field (a-c) and for different strengths
of the $\gamma_{\mathrm{rsb}}$ as a function of the subband index.
The error bars denote the standard deviation of the linear fit performed
for $\Delta V_{\mathrm{g}}(B)$ in Fig. \ref{fig:dftvgb}. The dashed
line on each plot indicates the nominal value of the $g=9$ used in
simulations. In case of $B=(0,0,B_{\mathrm{z}})$ we used data for
the range of the magnetic field for which $\Delta V_{\mathrm{g}}(B)$
had the linear behavior.}
\end{figure}

\section{Summary and Conclusions}
We described a numerical model to solve the transport problem for a biased InGaAs quantum
point contact for a two-dimensional electron gas with a confinement potential inducing narrowing
of the conducting channel. We studied the effects of the Rashba spin-orbit and electron-electron interaction
for transconductance in the regime of fractional conductance quantization in external magnetic field.
The calculated transconductance was used to extract the effective Land\'e factors as functions of
the external magnetic field orientation. For the extraction we simulated the standard experimental procedure
using the gate-voltage to energy conversion factor evaluated from the transconductance and its reaction to the source-drain bias.

We discussed the anisotropy of the $g^*$ factors looking for the effects due to the spin-orbit interaction and the interaction effects.
We have established that the spin-orbit interaction alone enhances the $g^*$ factor for the out-of-plane orientation of the magnetic
field. Moreover, the Rashba interaction introduces the in-plane anisotropy of the electron $g^*$ factors. The
$g^*$ factor for the magnetic field parallel to the current flow direction is unaffected by the SO interaction,
which reduces the $g^*$ factor for the perpendicular orientation of the in-plane magnetic field.
We explained that this effect -- counterintuitive from the point of view of the effective SO-related effective magnetic field --
is due to the subband mixing for the bottom of the subbands entering the transport window, for which non-zero transconductance is found.
The explanation was based on an analytical perturbation analysis.

The electron-electron interaction alone leads to a pronounced enhancement of the $g^*$ factor for out-of-plane magnetic field.
Moreover, the screening of the gate potential by the electron gas as described by the DFT approach induces asymmetry of the transconductance
lines on the gate potential - bias plane that is most pronounced near the pinch-off of the transport across the quantum point contact.
For combined electron-electron and Rashba interaction effects the calculated $g^*$ factors agree qualitatively with the experimental
data. We found that the electron-electron interaction preserves a weak dependence of the $g^*$ factor on the subbands index for the magnetic
field orientation parallel to the current flow, and the drop of the $g^*$ factor for the other perpendicular in-plane orientation of the magnetic field
vector.

\textbf{Acknowledgments} This work was supported by National Science
Centre according to decision DEC-2012/05/B/ST3/03290, and by PL-Grid
Infrastructure. The first author is supported by the scholarship of
Krakow Smoluchowski Scientific Consortium from the funding for National
Leading Reserch Centre by Ministry of Science and Higher Education
(Poland).

\section*{APPENDIX}

\subsection{Description of the scattering problem\label{sub:qtbm}}

In order to solve the scattering problem of the Fermi level electrons
inside the device discussed in this paper we implemented the QTBM
method in the finite difference formalism formally similar to the
tight-binding (TB) approach. The procedure developed here is a generalization
of our previous method \cite{kolasinski2014} with inclusion of the
spin-orbit coupling. We work with the wave-function description of
the transport with spinors component $\chi=\left(\begin{array}{c}
\chi^{\uparrow}\\
\chi^{\downarrow}
\end{array}\right)$ coupled by the SO interaction. The derivation is similar to the wave
function matching method \cite{Zwierzycki2008}. The general Schrödinger
equation for the system presented in Fig. \ref{fig:app1} can be written
in a compact form
\begin{equation}
\boldsymbol{\tau}_{n-1}\chi_{n-1}+\boldsymbol{H}_{n}\chi_{n}+\boldsymbol{\tau}_{n}^{\dagger}\chi_{n+1}=E_{\mathrm{F}}\chi_{n},\label{eq:appH}
\end{equation}
where $\boldsymbol{\tau}_{n\pm1}$ is the coupling matrix between
slice $n$ and $n\pm1$ and $\boldsymbol{H}_{n}$ is the TB Hamiltonian
describing $n$-th slice (see Fig. \ref{fig:app1}). 
The total wave function in the device contains contribution for all
the slices
\begin{equation}
\Psi_{\mathrm{D}}(u,v)=\sum_{k=1}^{\mathrm{N}}\chi_{k}(v)\delta_{u,k}.\label{eq:psiD}
\end{equation}

\begin{figure}[H]
\begin{centering}
\includegraphics[width=0.4\paperwidth]{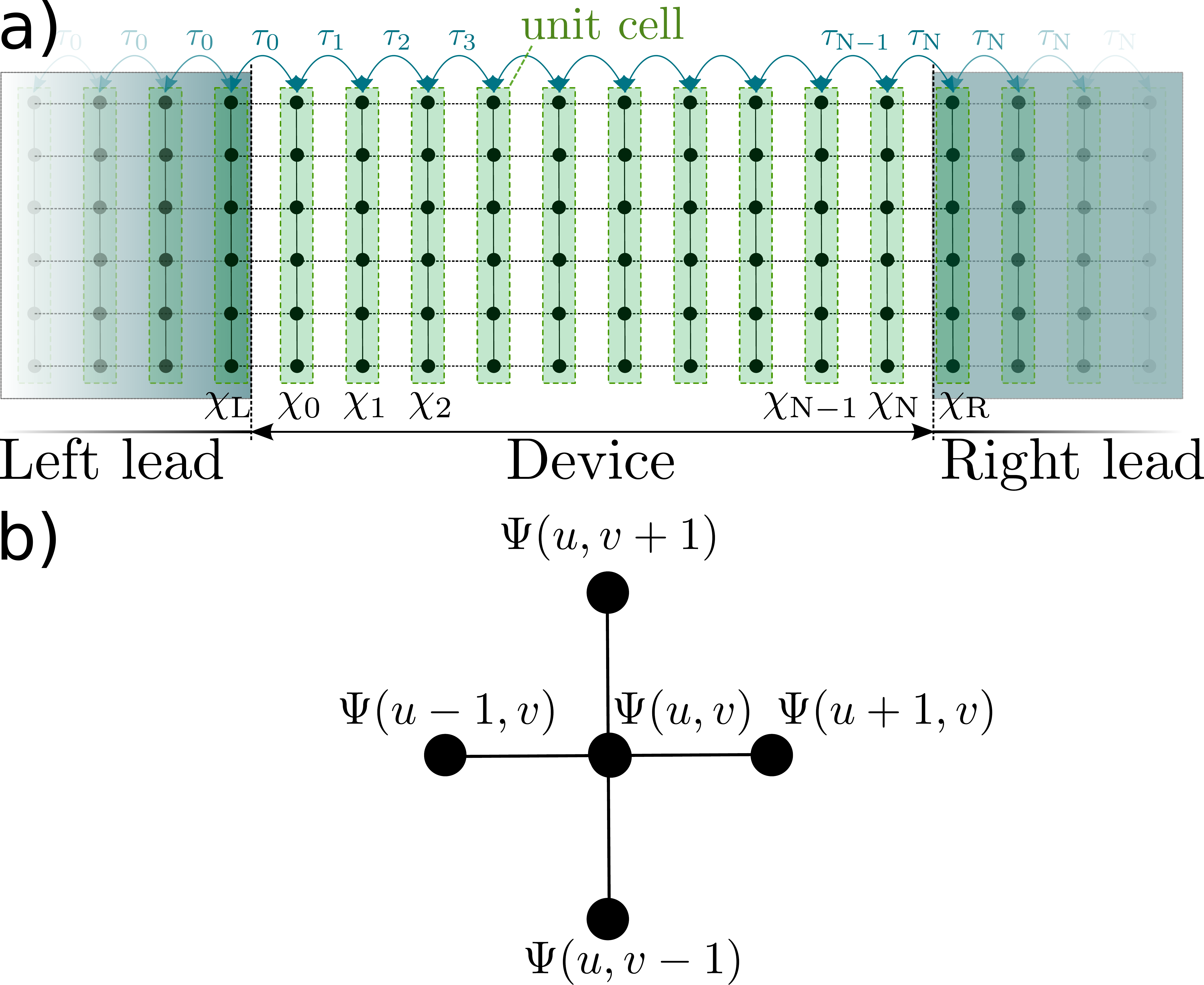}
\par\end{centering}

\caption{\label{fig:app1} (a) Schematic representation of the tight-binging
device considered in this paper. The system can be viewed as 1D chain
consistent of slices $\chi_{k}$ coupled by the matrix $\tau$. Here
$\chi$ is represented by the vector of size equal to number of points
in transverse direction. The left and right leads are assumed to be
ideal semi-infinite contacts. In our case points on the mesh are separated
by a distance $\Delta x$. The whole system consists of $N_{x}N_{y}$
grid points. (b) Coordinates of the scattering wave function on a
square TB mesh.}
\end{figure}

In order to find the scattering wave function one has to apply proper
boundary conditions at the device/lead interface which will satisfy
the continuity of the wave function and its derivative at the interface.
The standard approach for this problem is to find a general solution
in the homogeneous semi-infinite lead ($\tau_{n}=\tau_{0}$ for any
$n$ in the lead -- see Fig. \ref{fig:app1}). Thus the Eq. (\ref{eq:appH})
for the left lead can be written as
\begin{equation}
\boldsymbol{\tau}_{0}\chi_{n-1}+\boldsymbol{H}_{0}\chi_{n}+\boldsymbol{\tau}_{0}^{\dagger}\chi_{n+1}=E_{\mathrm{F}}\chi_{n}.\label{eq:appH2}
\end{equation}
This equation can be solved by applying Bloch substitution $\chi_{n}=\lambda^{n}\chi$
which assumes that the wave function between two nearest slices differs
by a constant factor $\lambda$. The number $\lambda$ can be interpreted
as phase which the wave function gains along a distance $\Delta x$
between two slices and can be written in a plane wave form $\lambda=e^{ik\Delta x}$.
After this substitution and division by $\lambda^{n-1}$ the Eq. (\ref{eq:appH2})
takes the form
\[
\boldsymbol{\tau}_{0}\chi+\boldsymbol{H}_{0}\lambda\chi+\boldsymbol{\tau}_{0}^{\dagger}\lambda^{2}\chi=E_{\mathrm{F}}\lambda\chi,
\]
which with substitution $\tilde{\chi}=\lambda\chi$ can be transformed
into the generalized eigenvalue problem for $\{\chi,\tilde{\chi}\}$
and eigenvalue $\lambda$
\begin{equation}
\left(\begin{array}{cc}
\boldsymbol{0} & \boldsymbol{I}\\
\boldsymbol{\tau}_{0} & \boldsymbol{H}_{0}-\boldsymbol{I}E_{\mathrm{F}}
\end{array}\right)\left(\begin{array}{c}
\chi\\
\tilde{\chi}
\end{array}\right)=\lambda\left(\begin{array}{cc}
\boldsymbol{I} & \boldsymbol{0}\\
\boldsymbol{0} & -\boldsymbol{\tau}_{0}^{\dagger}
\end{array}\right)\left(\begin{array}{c}
\chi\\
\tilde{\chi}
\end{array}\right).\label{eq:appEV}
\end{equation}
Upon solution of this eigenequation one obtains the transverse modes
$\chi_{m}$ in the lead and the corresponding wave vectors $k_{m}$.
We solve the eigenproblem (\ref{eq:appEV}) numerically with the LAPACK
library \cite{lapack}. 
In order to distinguish between the traveling and evanescent mode
we use fact that for the former $|\lambda_{m}|=1$ and for the later:
$|\lambda_{m}|>1$ is decaying reflected mode present in the left
lead near the scatterer -- and $|\lambda_{m}|<1$ is the decaying
transmission -- present in the right lead close to the scatterer.
We distinguish the traveling modes $\chi_{k}$ between the incoming
$\{\lambda_{k,+},\chi_{k,+}\}$ and outgoing $\{\lambda_{k,-},\chi_{k,-}\}$
ones modes by the sign of the carried current flux $\phi_{k}$: for
$\phi_{k}>0$ the mode was identified with $\chi_{k,+}$ and when
$\phi_{k}<0\Rightarrow\chi_{k,-}$. For the evanescent modes the flux
is always zero. The formula for $\phi_{k}$ can be obtained by considering
the continuity equation
\[
\frac{\partial\rho}{\partial t}=\boldsymbol{\nabla}\boldsymbol{j},
\]
in the TB formalism which leads to matrix formula for the current
flux $\phi_{k}\propto\Im\left(\lambda_{k}\chi_{k}^{\dagger}\tau^{\dagger}\chi_{k}\right)$\cite{Khomaykov2005}.
We denote $M$ as the number of the traveling modes in the lead. Note
that the number of the evanescent modes in general is infinite but
when we deal with TB like Hamiltonian this number is always finite
and equal to $M_{e}=N-M$, where $N$ is the size of the vector $\chi$.
Having sorted transverse modes one may write general solution for
the left semi-infinite lead at position $(u,v)$ on the mesh
\begin{equation}
\Psi_{\mathrm{L}}(u,v)=\sum_{k}^{M}a_{k,+}\lambda_{k,+}^{u}\chi_{k,+}+\sum_{k}^{M+M_{e}}r_{k,-}\lambda_{k,-}^{u}\chi_{k,-},\label{eq:psiL}
\end{equation}
where we assume that $\lambda_{k,-}^{u}\chi_{k,-}$ for $k\geq M+1$
correspond to decaying evanescent modes and $a_{k,+}$ is the scattering
amplitude of the incoming mode and $r_{k,-}$ is the reflection amplitude.
Analogical equation can be written for the right lead
\[
\Psi_{\mathrm{R}}(u,v)=\sum_{k}^{M}a_{k,-}\lambda_{k,-}^{u}\chi_{k,-}+\sum_{k}^{M+M_{e}}r_{k,+}\lambda_{k,+}^{u}\chi_{k,+}.
\]
We will denote by $\chi_{k,\pm}(v)$ the $v$ component of the $\chi_{k,\pm}$
vector. The idea of the QTBM is to match the wave functions inside
the leads $\Psi_{\mathrm{R}}$ and $\Psi_{\mathrm{L}}$ with the $\Psi_{\mathrm{D}}$.
The continuity condition at the left lead interface implies
\[
\Psi_{\mathrm{L}}(u=0,v)=\Psi_{\mathrm{D}}(u=0,v)=\chi_{0},\,\,\Psi_{u,v}\equiv\Psi_{\mathrm{D}}(u,v).
\]
Thus from Eq. (\ref{eq:psiL}) we have
\[
\Psi_{\mathrm{D}}(u=0,v)=\chi_{0}(v)=\sum_{k}^{M}a_{k,+}\chi_{k,+}+\sum_{k}^{M+M_{e}}r_{k,-}\chi_{k,-}.
\]
Following refs \cite{kolasinski2014,Leng1994,Kirkner1990,Luiser2014}
we express $r_{k,-}$ in terms of $\Psi_{\mathrm{L}}(0,v)$ and $a_{k,+}$
\begin{align}
r_{k,-} & =\sum_{p=1}^{M+M_{e}}\boldsymbol{S}_{kp,-}\left\langle \chi_{p,-}|\chi_{0}\right\rangle \nonumber \\
 & -\sum_{p=1}^{M+M_{e}}\sum_{q=1}^{M}\boldsymbol{S}_{kp,-}\boldsymbol{A}_{pq,-}a_{q,+},\label{eq:rk}
\end{align}
with $\left\langle \chi_{p,-}|\chi_{0}\right\rangle =\Delta x\sum_{v=1}^{N}\chi_{p,-}^{*}(v)\chi_{0}(v)$
being the inner product, $\boldsymbol{S}_{kp,-}^{-1}=\left\langle \chi_{k,-}|\chi_{p,-}\right\rangle $
and $\boldsymbol{A}_{kp,-}=\left\langle \chi_{k,-}|\chi_{p,+}\right\rangle $
are the transverse modes overlap matrices. Note that $\boldsymbol{S}_{-}$
is always a square matrix of dimensions $\left(M+M_{e},M+M_{e}\right)$,
while $\boldsymbol{A}_{-}$ has dimensions $\left(M+M_{e},M\right)$
and is a square matrix only when one neglects the evanescent modes
in the calculations.

The second condition that wave function at the interface lead/device
should fulfill is the continuity of its derivative at this point.
Numerically it can be written in the following way
\begin{align}
\Psi_{\mathrm{L}}(+1,v)-\Psi_{\mathrm{L}}(-1,v) & =\Psi_{+1,v}-\Psi_{-1,v}\label{eq:cderiv}\\
 & =\chi_{1}(v)-\chi_{-1}(v),
\end{align}
which is the finite difference equivalence for the analytical condition
\[
\left.\frac{\partial\Psi_{\mathrm{L}}}{\partial x}\right|_{x=0}=\left.\frac{\partial\Psi_{\mathrm{D}}}{\partial x}\right|_{x=0}.
\]
The left side of the Eq. (\ref{eq:cderiv}) can be evaluated using
Eq. (\ref{eq:psiL})
\begin{align*}
\Psi_{\mathrm{L}}(+1,v)-\Psi_{\mathrm{L}}(-1,v)= & \sum_{k}^{M}a_{k,+}\Delta_{k,+}\chi_{k,+}\\
+ & \sum_{k}^{M+M_{e}}r_{k,-}\Delta_{k,-}\chi_{k,-}\\
= & \chi_{1}(v)-\chi_{-1}(v)
\end{align*}
where $\Delta_{k,d}=\lambda_{k,d}-\lambda_{k,d}^{-1}$ and $d=(+,-)$.
Now inserting the expression for $r_{k,-}$ from Eq. (\ref{eq:rk})
to the equation above leads to
\begin{equation}
\chi_{1}(v)-\chi_{-1}(v)=\Gamma_{v}^{+}+\sum_{i=1}^{N}\boldsymbol{\Lambda}_{v,i}^{-}\chi_{0}(i),\label{eq:Psim1}
\end{equation}

with
\begin{align}
\Gamma_{v}^{+} & =\sum_{k}^{M}a_{k,+}\Delta_{k,+}\chi_{k,+}(v)-\sum_{k}^{M+M_{e}}\tilde{a}_{k,-}\Delta_{k,-}\chi_{k,-}(v)\label{eq:l1}\\
\tilde{a}_{k,-} & =\sum_{p=1}^{M+M_{e}}\sum_{q=1}^{M}\boldsymbol{S}_{kp,-}\boldsymbol{A}_{pq,-}a_{q,+}\label{eq:l2}\\
\boldsymbol{\Lambda}_{v,i}^{-} & =\Delta x\sum_{k=1}^{M+M_{e}}\sum_{p=1}^{M+M_{e}}\Delta_{k,-}\chi_{k,-}(v)\boldsymbol{S}_{kp,-}\chi_{p,-}^{*}(i).\label{eq:l3}
\end{align}
With Eq. (\ref{eq:Psim1}) we can calculate the value of the wave
function outside the device $\chi_{-1}$. Using the vector notation
(see Fig. \ref{fig:app1} and Eq. (\ref{eq:appH})) we get following
formula
\[
\chi_{-1}=\chi_{1}-\Gamma^{+}-\boldsymbol{\Lambda}^{-}\chi_{0},
\]
which we put to the Eq. (\ref{eq:appH}) for $n=0$ in order to remove
the reference to the slice $\chi_{-1}$ outside the device
\[
\boldsymbol{\tau}_{-1}\left(\chi_{1}-\Gamma^{+}-\boldsymbol{\Lambda}^{-}\chi_{0}\right)+\boldsymbol{H}_{0}\chi_{0}+\boldsymbol{\tau}_{0}^{\dagger}\chi_{1}=E_{\mathrm{F}}\chi_{0},
\]
where we assume, for simplicity that lead are far from the scattering
center, which is usually located in the center of the device, thus
we can put $\boldsymbol{\tau}_{-1}=\boldsymbol{\tau}_{0}$. Simplifying
the equation above we get expression for the boundary condition in
the left lead
\[
\left(\boldsymbol{H}_{0}-\boldsymbol{I}E_{\mathrm{F}}-\boldsymbol{\tau}_{0}\boldsymbol{\Lambda}^{-}\right)\chi_{0}+\left(\boldsymbol{\tau}_{0}+\boldsymbol{\tau}_{0}^{\dagger}\right)\chi_{1}=\boldsymbol{\tau}_{0}\Gamma^{+}.
\]
Analogically to Eq. (\ref{eq:Psim1}) we get similar equation for
the right lead boundary condition
\[
\Psi_{\mathrm{N}+1,v}-\Psi_{\mathrm{N}-1,v}=\Gamma_{v}^{-}+\sum_{i=1}^{N}\boldsymbol{\Lambda}_{v,i}^{+}\Psi_{\mathrm{N},v},
\]
where the definitions for vector $\Gamma_{v}^{-}$ and matrix $\boldsymbol{\Lambda}_{v,i}^{+}$
can be obtained from equations (\ref{eq:l1}-\ref{eq:l3}) by changing
sign $+\rightarrow-$ and $-\rightarrow+$, and noting that the calculations
of those quantities are performed for $N$-th slice of the device.
From the equation above we calculate the formula for $\mathrm{N+1}$
slice using the vector notation
\[
\chi_{\mathrm{N}+1}=\chi_{\mathrm{N}-1}+\Gamma^{-}+\boldsymbol{\Lambda}^{+}\chi_{\mathrm{N}},
\]
and we substitute it to the Eq. (\ref{eq:appH}) for $n=N$ assuming
that $\boldsymbol{\tau}_{\mathrm{N}}=\boldsymbol{\tau}_{\mathrm{N-1}}$,
\[
\boldsymbol{\tau}_{\mathrm{N-1}}\chi_{\mathrm{N-1}}+\boldsymbol{H}_{\mathrm{N}}\chi_{\mathrm{N}}+\boldsymbol{\tau}_{\mathrm{N}}^{\dagger}\left(\chi_{\mathrm{N-1}}+\Gamma^{-}+\boldsymbol{\Lambda}^{+}\chi_{\mathrm{N}}\right)=E_{\mathrm{F}}\chi_{\mathrm{N}},
\]
which leads to final formula for the right lead
\[
\left(\boldsymbol{\tau}_{\mathrm{N}}+\boldsymbol{\tau}_{\mathrm{N}}^{\dagger}\right)\chi_{\mathrm{N-1}}+\left(\boldsymbol{H}_{\mathrm{N}}-\boldsymbol{I}E_{\mathrm{F}}+\boldsymbol{\tau}_{\mathrm{N}}^{\dagger}\boldsymbol{\Lambda}^{+}\right)\chi_{\mathrm{N}}=-\boldsymbol{\tau}_{\mathrm{N}}^{\dagger}\Gamma^{-}.
\]
Finally the whole matrix equation can be written in the form of the
system of linear equations $\boldsymbol{A}\chi=b$, with block matrices
{\scriptsize{{{
\begin{equation}
\boldsymbol{A}=\left(\begin{array}{ccccc}
\boldsymbol{H}'_{0}-\boldsymbol{\tau}_{0}\boldsymbol{\Lambda}^{-} & \boldsymbol{\tau}_{0}+\boldsymbol{\tau}_{0}^{\dagger} &  & \cdots & 0\\
\boldsymbol{\tau}_{0} & \boldsymbol{H}'_{1} & \boldsymbol{\tau}_{1}^{\dagger} &  & \vdots\\
 & \ddots & \ddots & \ddots\\
\vdots &  & \boldsymbol{\tau}_{\mathrm{N-2}}^{\dagger} & \boldsymbol{H}'_{N-1} & \boldsymbol{\tau}_{\mathrm{N-1}}^{\dagger}\\
0 & \cdots &  & \boldsymbol{\tau}_{\mathrm{N}}+\boldsymbol{\tau}_{\mathrm{N}}^{\dagger} & \boldsymbol{H}'_{\mathrm{N}}+\boldsymbol{\tau}_{\mathrm{N}}^{\dagger}\boldsymbol{\Lambda}^{+}
\end{array}\right),\label{eq:mA}
\end{equation}
}}}}with $\boldsymbol{H}'_{n}=\boldsymbol{H}{}_{n}-\boldsymbol{I}E_{\mathrm{F}}$,
$\chi=(\chi_{0},\chi_{1},\ldots,\chi_{\mathrm{N-1}},\chi_{\mathrm{N}})$
and $b=(\boldsymbol{\tau}_{0}\Gamma^{+},0,\ldots,0,-\boldsymbol{\tau}_{\mathrm{N}}^{\dagger}\Gamma^{-})$.
In order to obtain the scattering wave function for the $p$-th mode
incoming to device from the left one must put $a_{k,+}=\delta_{k,p}$
in the definition of the $\Gamma^{+}$ and $a_{k,-}=0$ for all $k$
in the $\Gamma^{-}$. For the current flow from right to left we put
$a_{k,-}=\delta_{k,p}$ in $\Gamma^{-}$ and $a_{k,+}=0$ in the definition
of the $\Gamma^{+}$. We use the PARDISO library to solve the system of
linear equations \cite{PARDISO}.

Let us denote the k-th scattering mode by $\Psi^{k}$ where $k=(1,2,...,M)$
enumerates modes in the left lead. Once the system of linear equations
is solved one may calculate the reflection amplitudes from Eq. (\ref{eq:rk})
by calculating the value of $r_{k}\equiv r_{k,-}$, and transmission
amplitudes $t_{k}\equiv r_{k,+}$, which in case of the transverse
mode incoming from the left lead we have
\[
t_{k}=r_{k,+}=\sum_{p=1}^{M+M_{e}}\boldsymbol{S}_{kp,+}\left\langle \chi_{p,+}|\Psi_{\mathrm{N},v}^{k}\right\rangle .
\]
Using calculated values of $r_{k}$ and $t_{k}$ we calculate the
reflection $R_{p}$ and transmission $T_{p}$ probability of the $p$-th
incoming mode from the left lead with equation
\begin{align*}
R_{p} & =\sum_{k=1}^{M}\left|r_{k}\right|^{2}\left|\frac{\phi_{\mathrm{L},k,-}}{\phi_{\mathrm{L},p,+}}\right|,\\
T_{p} & =\sum_{k=1}^{M}\left|t_{k}\right|^{2}\left|\frac{\phi_{\mathrm{R},k,+}}{\phi_{\mathrm{L},p,+}}\right|,
\end{align*}
where $\phi_{\{L,P\},k,\{+,-\}}$ denotes the quantum flux which mode
number $k$ is carrying in the $\mathrm{L}$ - left lead or $\mathrm{R}$
- right lead in the $+$- right direction or -- left direction (outgoing
modes in the left lead). From that we calculate the conductance of
the system using the Landauer formula
\[
G=\frac{e^{2}}{h}\sum_{p=1}^{M}T_{p}.
\]

The tight-binding form of the Schrödinger equation was obtained by
applying the gauge-invariant kinetic-energy finite difference discretization
\cite{governale1998} of the Hamiltonian given by Eq. (\ref{ham}).
This discretization leads to following matrix equation for the device
wave function
\begin{equation}
\left(\begin{array}{cc}
\chi_{\mathrm{uv}}^{\uparrow}D_{\mathrm{uv}}^{\uparrow}+\Delta_{\mathrm{uv}}^{\uparrow} & \chi_{\mathrm{uv}}^{\downarrow}S_{b}^{\downarrow}+\Gamma_{\mathrm{uv}}^{\downarrow}\\
\chi_{\mathrm{uv}}^{\uparrow}S_{b}^{\uparrow}+\Gamma_{\mathrm{uv}}^{\uparrow} & \chi_{\mathrm{uv}}^{\downarrow}D_{\mathrm{uv}}^{\downarrow}+\Delta_{\mathrm{uv}}^{\downarrow}
\end{array}\right)=E_{\mathrm{F}}\left(\begin{array}{c}
\chi_{\mathrm{uv}}^{\uparrow}\\
\chi_{\mathrm{uv}}^{\downarrow}
\end{array}\right),\label{eq:fdham}
\end{equation}

with
\begin{align*}
D_{\mathrm{uv}}^{\mathrm{\sigma}} & =\frac{2}{\Delta x^{2}}+eV_{\mathrm{ext}}(u,v)+\frac{1}{2}\sigma g\mu_{\mathrm{B}}B_{\mathrm{z}},\\
T_{\mathrm{uv}}^{\mathrm{d\sigma}} & =-\frac{1}{2\Delta x^{2}}e^{+id\sigma\gamma_{\mathrm{lat}}E_{\mathrm{x}}\left(u,v\right)\Delta x},\\
S^{\mathrm{d}} & =-d\frac{i\gamma_{\mathrm{rsb}}}{2\Delta x},\\
S_{\mathrm{v}}^{\mathrm{d\sigma}} & =-ds\frac{i\gamma_{\mathrm{rsb}}}{2\Delta x}e^{-di\Delta x^{2}B_{\mathrm{z}}v},\\
S_{\mathrm{b}}^{\sigma} & =\frac{1}{2}g\mu_{\mathrm{B}}\left(B_{\mathrm{x}}+siB_{\mathrm{y}}\right),\\
\Delta_{\mathrm{uv}}^{\sigma} & =\chi_{\mathrm{uv-1}}^{\sigma}T_{uv-1}^{-\sigma}+\chi_{\mathrm{uv+1}}^{\sigma}T_{uv}^{+\sigma}+\chi_{\mathrm{u-1v}}^{\sigma}\tau_{u-1v}^{-\sigma}\\
 & +\chi_{\mathrm{u+1v}}^{\sigma}\tau_{uv}^{+\sigma},\\
\Gamma_{\mathrm{uv}}^{\sigma} & =\chi_{\mathrm{u-1v}}^{\sigma}S_{\mathrm{v}}^{\mathrm{-\sigma}}+\chi_{\mathrm{u+1v}}^{\sigma}S_{\mathrm{v}}^{\mathrm{+\sigma}}+\chi_{\mathrm{uv-1}}^{\sigma}S^{\mathrm{-}}\\
 & +\chi_{\mathrm{uv+1}}^{\sigma}S^{\mathrm{+}},\\
\tau_{\mathrm{uv}}^{\mathrm{d\sigma}} & =-\frac{1}{2\Delta x^{2}}e^{-di\Delta x^{2}B_{\mathrm{z}}v-id\sigma\gamma_{\mathrm{lat}}E_{\mathrm{y}}\left(u,v\right)\Delta x},
\end{align*}
where $d=\pm$, and the electric fields generated by the in-plane
external potentials $E_{\mathrm{x}}(u,v)=\left(V_{\mathrm{ext}}(u+1,v)-V_{\mathrm{ext}}(u-1,v)\right)/2\Delta x$
and $E_{\mathrm{y}}(u,v)=\left(V_{\mathrm{ext}}(u,v+1)-V_{\mathrm{ext}}(u,v-1)\right)/2\Delta x$.

We have mapped the 2D+spin coordinates $\chi_{uv}^{\sigma}$ into
1D coordinate using following formula: $w(u,v,\sigma)=\left(\frac{\sigma+3}{2}\right)\left(uN_{y}+v\right)$
in order to get row or column index of the TB matrix. Note that such
parametrization of $w$ divides the entire matrix into two major blocks
\[
\left(\begin{array}{cc}
\boldsymbol{H}^{\uparrow} & \boldsymbol{S}^{\uparrow\downarrow}\\
\boldsymbol{S}^{\downarrow\uparrow} & \boldsymbol{H}^{\downarrow}
\end{array}\right)\left(\begin{array}{c}
\chi^{\uparrow}\\
\chi^{\downarrow}
\end{array}\right)=E_{\mathrm{F}}\left(\begin{array}{c}
\chi^{\uparrow}\\
\chi^{\downarrow}
\end{array}\right),
\]
with $\boldsymbol{H}^{\sigma}$ being the Hamiltonian of the upper
or lower part of the spinor and $\boldsymbol{S}^{\sigma\sigma'}$
being the coupling matrices. When there is no SO interaction or magnetic
field is oriented only in $z$ direction the matrix can be easily
divided into two independent subspaces, twice times smaller. In order
to calculate the conductance we applied the QTBM described above to
Hamiltonian Eq. (\ref{eq:fdham}). The dispersion relations presented
in Fig. \ref{fig:reldysp} were obtained by substituting the plane
wave of form
\[
\chi_{\mathrm{uv}}=e^{iku\Delta x}\left(\begin{array}{c}
\chi_{v}^{\uparrow}\\
\chi_{v}^{\downarrow}
\end{array}\right),
\]
into Eq. (\ref{eq:fdham}) and then calculating the eigenvalues of
created eigenproblem as a function of $k$.

\subsection{Description of the many electron problem}

We solve the many electron problem using the Density Functional Theory
(DFT) with Local Spin-Density Approximation (LSDA) for the evaluation
of the exchange-correlation functional. The approach used in this
paper is similar to those described in our previous work \cite{kolasinskiDFT2013}
or in \cite{szafranDFT2011}. All the DFT calculations were done for
$V_{\mathrm{SD}}=0$ case. The Kohn-Sham (KS) orbitals were obtained
by diagonalizing the Hamiltonian given by Eq. (\ref{eq:fdham}) with
the spin dependent exchange-correlation potential $V_{\mathrm{XC}}^{\mathrm{s}}$
and Hartree $V_{\mathrm{H}}$ potential added to the diagonal terms
\begin{align}
D_{\mathrm{uv}}^{\mathrm{s}} & =\frac{2}{\Delta x^{2}}+eV_{\mathrm{ext}}(u,v)+\frac{1}{2}sg\mu_{\mathrm{B}}B_{\mathrm{z}}\nonumber \\
 & +eV_{\mathrm{H}}(u,v)+eV_{\mathrm{XC}}^{\mathrm{s}}(u,v).\label{eq:dftham}
\end{align}
The periodic boundary conditions were applied in the $x$ direction
in order to simulate open infinite leads. We use the Attaccalite parametrization
for the $V_{\mathrm{XC}}^{\mathrm{s}}$ potential \cite{attaccalite2002},
which depends on the spin-up electron density $n_{\uparrow}$ and
spin-down $n_{\downarrow}$. The Hartree potential was calculated
by numerical integration of the Coulomb interaction term
\[
V_{\mathrm{H}}(x,y)=\frac{e}{4\pi\epsilon\epsilon_{0}}\left(-I_{\mathrm{d}}(x,y;20\mathrm{nm})+I_{\mathrm{e}}(x,y;0)\right),
\]
where we used $\epsilon=12.4$ for InGaAs, and $I_{\mathrm{d}}$ is
the coulomb integral over the positively charged donor layer and $I_{e}$
-- integral over the total electron density in 2DEG layer and
\[
I_{\mathrm{x}}(x,y;z_{\mathrm{x}})=\int dx'dy'\frac{n_{\mathrm{x}}(x',y')}{\left|(x,y,0)-(x',y',z_{\mathrm{x}})\right|}.
\]
In order to simulate the infinite leads we performed the integration
including 8 copies of the system in left and right direction. In the
formula above we assume that the 2D donor layer is located $z_{\mathrm{d}}=20$nm
above the layer of 2DEG ($z_{\mathrm{e}}=0$nm) and the density of
the ionized donors in that layer is constant. The total number of
donors was $N_{\mathrm{d}}=150$ in our calculations and the total
electron density was calculated as $n_{\mathrm{e}}(x,y)=n_{\uparrow}(x,y)+n_{\downarrow}(x,y)$.
In the first step of DFT iterative scheme we assumed that $n_{\mathrm{e}}=n_{\uparrow}=n_{\downarrow}=0$,
then we performed the numerical diagonalization of Hamiltonian (\ref{eq:fdham},\ref{eq:dftham})
with the FEAST eigenvalue solver \cite{feast2009}, calculating the
first lowest $M_{\mathrm{e}}=500$ eigen-energies $\varepsilon_{i}^{\sigma}$
and corresponding KS spin-orbitals $\chi_{i}^{\sigma}\equiv\chi_{i}^{\sigma}(u,v)$.
From that we have calculated the electron densities
\begin{equation}
n_{\sigma}\equiv n_{\sigma}(u,v)=\sum_{k=1}^{M_{\mathrm{e}}}f(\varepsilon_{k}^{\sigma};E_{\mathrm{F}},T)\left|\chi_{k}^{\sigma}\right|^{2},\label{eq:dftn}
\end{equation}
where $f$ is the Fermi-Dirac distribution
\[
f(\varepsilon_{k}^{\sigma};E_{\mathrm{F}},T)=\frac{1}{e^{\left(\varepsilon_{k}^{\sigma}-E_{\mathrm{F}}\right)/k_{\mathrm{B}}T}+1},
\]
and the Fermi energy $E_{\mathrm{F}}$ was obtained from the charge
neutrality condition
\begin{equation}
N_{\mathrm{d}}=\sum_{k=1}^{M_{\mathrm{e}}}f(\varepsilon_{k}^{\uparrow};E_{\mathrm{F}},T)+f(\varepsilon_{k}^{\downarrow};E_{\mathrm{F}},T).\label{eq:dftN}
\end{equation}
We solved both Eq. (\ref{eq:dftn}) and (\ref{eq:dftN}) for densities
$n_{\sigma}$ and $E_{\mathrm{F}}$ using the bisection method. Once
the $n_{\mathrm{e}},n_{\uparrow}$ and $n_{\downarrow}$ were obtained
we have calculated the $V_{\mathrm{H}}$ and $V_{\mathrm{XC}}^{\mathrm{s}}$
potentials then we solved the KS Eq. (\ref{eq:fdham},\ref{eq:dftham})
again. The whole procedure was repeated until the density stopped
to change significantly between two consecutive iterations.

In order to improve the convergence of the DFT iterative scheme we
have implemented a) the temperature annealing b) density mixing method.

In case of temperature annealing we start the calculation with temperature
$T=$10K, then after the convergence was obtained we use calculated
electron densities as input densities for lower temperature calculations.
The annealing was performed in three steps for $T=\{10K,4K,0.5K\}$
temperatures.

In case of density mixing we applied the simple local and linear mixing
scheme \cite{Certik}. Let us denote by $n_{\sigma}^{k}$ the electron
density with spin $\sigma$ in $k$-th DFT iteration, and by $n_{\sigma}^{k+1}\equiv F(n_{\sigma}^{k})$
electron density obtained after diagonalization process from Eq. (\ref{eq:dftn}).
If we define the residual vector $R(n_{\sigma}^{k})\equiv F(n_{\sigma}^{k})-n_{\sigma}^{k}$
then standard approach for almost all mixing schemes start from equation
\begin{equation}
n_{\sigma}^{k+1}:=n_{\sigma}^{k}+\boldsymbol{J}^{-1}R(n_{\sigma}^{k}),\label{eq:mixing}
\end{equation}
where $\boldsymbol{J}_{ij}^{\sigma}=\frac{\partial R_{i}(n_{\sigma}^{k})}{\partial x_{j}}$
is the Jacobian matrix, $R_{i}(n_{\sigma}^{k})$ is the $i$-th component
of the residual vector (the residuum value for $i$-th point in space)
and $:=$ represents the numerical substitution. In the simplest case
one may assume that Jacobian is a constant diagonal matrix $\boldsymbol{J}^{\sigma}=-\alpha\boldsymbol{I}$,
where $0<\alpha<1$ is the mixing parameter, chosen arbitrarily for
a given problem, then the mixing formula (\ref{eq:mixing}) for electron
density in the next steps takes the from
\begin{equation}
n_{\sigma}^{k+1}:=n_{\sigma}^{k}+\alpha\left(F(n_{\sigma}^{k})-n_{\sigma}^{k}\right)=\alpha n_{\sigma}^{k+1}+\left(1-\alpha\right)n_{\sigma}^{k}.\label{eq:sm}
\end{equation}
In the second approximation one may assume that Jacobian is a diagonal
matrix, which change during the iteration, of form
\begin{equation}
\left[\boldsymbol{J}_{k}^{\sigma}\right]^{-1}=-\mathrm{diag}(\alpha_{1}^{\sigma},\alpha_{2}^{\sigma},\ldots,\alpha_{N}^{\sigma}),\label{eq:jac}
\end{equation}
where $N$ is the size of the vector $R$. This method start with
$\alpha_{1}^{\sigma}=\alpha_{2}^{\sigma}=\ldots=\alpha_{N}^{\sigma}=\alpha_{\mathrm{min}}$
then at each iteration the parameters $\alpha_{i}^{\sigma}$ are changed
according the following algorithm: if $R_{i}(n_{\sigma}^{k})R_{i}(n_{\sigma}^{k-1})>0$
then increase the value of $\alpha_{i}^{\sigma}$ by $\alpha_{\mathrm{min}}$
(if $\alpha_{i}^{\sigma}>\alpha_{\mathrm{max}}$ then $\alpha_{i}^{\sigma}=\alpha_{\mathrm{max}}$)
otherwise set $\alpha_{i}^{\sigma}=\alpha_{\mathrm{min}}$. Note that
the method works for $k\geq3$, hence for $k=1$ we use the direct
substitution $n_{\sigma}^{k+1}:=F(n_{\sigma}^{k})$. In second iteration
$k=2$ we used the simple mixing formula (\ref{eq:sm}) with $\alpha=0.05$.
Then for $k\geq3$ we implemented the local mixing algorithm given
by Jacobian (\ref{eq:jac}), with $\alpha_{\mathrm{min}}=0.05$ and
$\alpha_{\mathrm{max}}=0.8$. According to the paper \cite{Certik}
this method is almost as efficient as the more complicated Broyden
method \cite{broyden1965}.

Once the convergence of the KS equations was obtained we use the obtained
from Eq. (\ref{eq:dftn}) and (\ref{eq:dftN}) Fermi energy $E_{\mathrm{F}}$
and calculated $V_{\mathrm{H}}$ and $V_{\mathrm{XC}}^{\mathrm{s}}$
potentials in the one electron picture method described in section
\ref{sub:qtbm}.

\bibliographystyle{apsrev4-1}

\bibliography{sosok7}

\end{document}